\documentclass[11pt,a4paper]{article}

\usepackage{geometry}
\usepackage{natbib}

\usepackage{graphicx,amsmath}
\usepackage{epstopdf,epsfig}

\usepackage{hyperref}
\hypersetup{
 hidelinks,
}

\usepackage{color}

\usepackage{lipsum}

\usepackage{siunitx}

\renewcommand\Re{\mbox{Re}}

\newcommand\Ka{\mbox{Ka}}
\newcommand\Da{\mbox{Da}}
\newcommand\Le{\mbox{Le}}

\newcommand\ssKa{\mbox{\scriptsize{Ka}}}

\newcommand{\PI}{\Pi_\varepsilon}

\newif\iflowres
\lowresfalse

\newif\ifshortsolver
\shortsolverfalse
\shortsolvertrue

\title{Towards the Distributed Burning Regime in\\Turbulent Premixed Flames}

\author{A. J. Aspden$^{1,2}$, M. S. Day$^2$ and J. B. Bell$^2$}

\date{{\small{$^1$School of Engineering, Newcastle University, Stephenson Building,\\
Claremont Road, Newcastle-Upon-Tyne, NE1 7RU, UK\vspace{2mm}\\
$^2$Center for Computational Sciences and Engineering, Lawrence Berkeley National Laboratory,\vspace{-1mm}\\
MS50A-3100, 1 Cyclotron Road, Berkeley, CA 94720, USA}}\vspace{4mm}\\
25th June 2018}

\begin{document}

\maketitle

\begin{abstract}
Three-dimensional numerical simulations of canonical statistically-steady statistically-planar
turbulent flames have been used in an attempt to produce distributed burning in lean methane and 
hydrogen flames.  Dilatation across the flame means that extremely large Karlovitz numbers are 
required; even at the extreme levels of turbulence studied (up to a Karlovitz number of 8767) 
distributed burning was only achieved in the hydrogen case.  In this case, turbulence 
was found to broaden the reaction zone visually by around an order of magnitude, and thermodiffusive 
effects (typically present for lean hydrogen flames) were not observed.  In the preheat zone, the 
species compositions differ considerably from those of one-dimensional flames based a number of 
different transport models (mixture-averaged, unity Lewis number, and a turbulent eddy viscosity 
model).  The behaviour is a characteristic of turbulence dominating non-unity Lewis number species 
transport, and the distinct limit is again attributed to dilatation and its effect on the 
turbulence.  Peak local 
reaction rates are found to be lower in the distributed case than in the lower Karlovitz cases but 
higher than in the laminar flame, which is attributed to effects that arise from the modified 
fuel-temperature distribution that results from turbulent mixing dominating low Lewis number 
thermodiffusive effects.  Finally, approaches to achieve distributed burning at 
realisable conditions are discussed; factors that increase the likelihood of
realising distributed burning are higher pressure, lower equivalence ratio, higher Lewis number,
and lower reactant temperature.
\end{abstract}

\section{Introduction}

The distributed burning regime of turbulent premixed flames represents the limiting case where flame
propagation is driven by turbulent mixing rather than molecular diffusion \citep{AspdenJFM11}, 
and corresponds to the small-scale turbulence limit \citep{Peters00,Damkohler40};
the Reader is referred to \citet{AspdenJFM11} for a more detailed review of the use of the term 
distributed burning.
The feature that distinguishes burning in the distributed mode is that turbulent eddies
comparable with the reaction zone thickness can mix faster than the flame can burn.
Experimentally, there is some evidence of what will be referred to here as 
``transitionally-distributed''
burning, particularly in high-speed piloted lean-to-stoichiometric methane jet flames; see, for 
example, \citet{Dunn10}, \citet{Zhou17}, \citet{SkibaCnF18}, and the references therein.
Previous numerical studies of transitionally-distributed flames include \citet{PoludnenkoCnF10}, 
\citet{AspdenJFM11}, \citet{SavardCnF15}, \citet{LapointeCnF15}, \citet{NilssonFuel18},
\citet{WangCnF18}, and the references therein.
All of these studies have demonstrated flames with broadened preheat zones,
but none have shown broadening of the actual reaction zone by turbulence alone.  
\cite{SkibaCnF18} referred to this mode of burning as ``broadened preheat thin reactions''.

Distributed burning has been observed numerically in an astrophysical context \citep{Aspden08a},
where a single-step reaction model was used to represent thermonuclear fusion in a type Ia
supernova flame.
Subsequently, \citet{Aspden10} demonstrated scaling laws for distributed supernova flames
following \citet{Damkohler40} and \cite{Peters00}, along with the so-called ``$\lambda$-flame''
regime; the combination of large Karlovitz and Damk\"ohler numbers gives rise to flames
simultaneously in the small-scale and large-scale limits.  The resulting regime has such a large
range of turbulent scales that the flame burns in the distributed mode, but the larger scales
are unable to mix before the flame burns; this mode of burning aligns with that predicted by
\cite{Zimont79} (which we argue requires the distributed transition to broaden the flame),
and the local turbulent flame thickness $\lambda$ corresponds to the Zimont length
\citep{Peters00} at that Karlovitz number.
These supernova studies used an idealised configuration capable of subjecting the flame
to arbitrary levels of turbulence favourable for distributed burning (the Reynolds number in a
supernova can be in excess of $10^{10}$ and the Mach number around $10^{-5}$); realisable
conditions for distributed burning in terrestrial chemical flames are yet to be established.

In the present paper, lean premixed methane and hydrogen flames have been simulated with extreme
levels of turbulence (rms velocity fluctuations exceeding four hundred times the laminar flame
speed).  At the highest turbulence levels, the hydrogen flame has been found to present
substantial broadening of the reaction zone, but even with such intense turbulence, the methane
flame did not.  Despite the abstracted configuration and unrealistic conditions, this hydrogen
flame represents the transition to distributed burning expected of a terrestrial chemical flame
if suitable conditions can be contrived (i.e.~at the same Karlovitz and Damk\"ohler numbers).  
Phenomenological observations are first presented, followed by consideration of global
consumption speeds, flame thickening, and conditional means of heat release and species mass
fractions.  The paper concludes with a discussion of the distributed burning regime and
potential conditions required to realise distributed burning at realisable conditions.

\subsection{A note on Karlovitz number}

The classical Richardson/Kolmogorov picture of turbulence is a cascade of turbulent eddies 
from the energy-containing large scales through the inertial subrange to the dissipation subrange. 
If it is assumed that turbulence is sufficient for the inertial subrange to extend down to the
flame scale, (equivalently, the flame is in the thin reaction zone, or the Karlovitz number is
larger than unity), then as in the classical picture, there are energy-containing large scales
responsible for the supply of kinetic energy through the inertial subrange to the flame scale,
where (unlike the non-reacting case) dilatation modifies the turbulent structure close to the
flame before it reaches a classical dissipation subrange.
Energy is still dissipated by viscosity at small scales, but not at 
the usual Kolmogorov length scale that would be expected in a constant density flow;
see \cite{ToweryPRE2016} for a more detailed consideration of energy spectra in premixed flames.

For this reason, it seems illogical to define the Karlovitz number (in the sense of being 
characteristic of turbulence-flame interaction at the flame scale) in terms of Kolmogorov scales;
these scales simply aren't representative of the physical processes taking place.
Following the Kolmogorov similarity hypotheses, the universal equilibrium range is determined by
the energy dissipation rate $(\varepsilon)$ and viscosity ($\nu$).  
The second similarity hypothesis gives rise to the inertial subrange independent from $\nu$, 
determined solely by $\varepsilon$.
This second hypothesis appears to be relevant to a turbulent premixed flame, which suggests a 
relevant dimensionless parameter can be defined as
\begin{equation}
\PI^2=\frac{\varepsilon}{\varepsilon_F}=\frac{{u^\prime}^3}{l}\frac{l_F}{s_F^3},
\label{eq:Ka}
\end{equation}
where $u^\prime$ and $l$ are the rms velocity fluctuation and integral length scale respectively,
and $s_F$ and $l_F$ are the characteristic speed and thermal thickness of the flame 
(and $\varepsilon_F=s_F^3/l_F$).
This quantity is the same as a more conventionally-defined Karlovitz number based on the 
Kolmogorov time scale and by assuming a unity flame Reynolds
number i.e.~$s_Fl_F=\nu$ \citep[e.g.][]{Peters00}, but
the argument here is that (\ref{eq:Ka}) is the appropriate dimensionless parameter that
characterises the strength of turbulence at the flame scale in the thin reaction zone and above,
(specifically, $u_F/s_F=\Pi_\varepsilon^{2/3}$, where $u_F$ is the velocity associated with a
turbulent eddy in the inertial subrange at the flame scale),
and so should be considered the starting point, rather than resulting from any assumption.
It is not only important to distinguish this quantity from the conventional definition involving
Kolmogorov scales, but also from the definition used in the flamelet regimes as a measure of
flame stretch; these definitions represent fundamentally different physical interactions.
Despite the distinction, $\PI$ will still be referred to here as a Karlovitz number
(with $\Ka\equiv\PI$) and the implicit understanding that it is not a measure of stretch, 
nor does it involve viscosity.

\section{Simulation Details}

\subsection{Numerical solver}

The numerical solver used here is based on the well-established low Mach number
formulation of the reacting flow equations \citep{DayBell2000,Nonaka2012}.
The fluid is treated as a mixture of perfect gases, and 
a mixture-averaged model is assumed for diffusive transport.
A source term is used in the momentum equation
to establish and maintain turbulence with the desired properties \citep{Aspden08b}.
The chemical kinetics and transport are modelled using 
the hydrogen mechanism of \citet{LiDryer2004}
consisting of 9 species with 21 fundamental reactions,
and the GRIMech 3.0 methane mechanism \citep{FrenklachWang1995} 
with the nitrogen reactions removed, resulting in 35 species and 217 reactions.
These evolution equations are supplemented by
CHEMKIN-compatible databases for thermodynamic quantities, 
and transport properties computed using EGLIB \citep{EGLIB}.

This solver is capable of running implicit large eddy simulation (ILES);
non-oscillatory finite-volume schemes such as this are able to dissipate kinetic energy
numerically at the grid scale in a stable and physical manner without resolving all the
way down to the Kolmogorov length scale (see \cite{GrinsteinBook07} for a review).  
In the present simulations, especially at the highest turbulence intensities, there
is some reliance on this ILES capability in that the Kolmgorov length scale in the reactants
is not resolved on the grid.  The performance of this solver in such under-resolved conditions
was characterised in \citet{Aspden08b}, and further details are given below.

\subsection{Simulation configuration}

\begin{figure}[t!]
\centering
\includegraphics[width=0.48\textwidth]{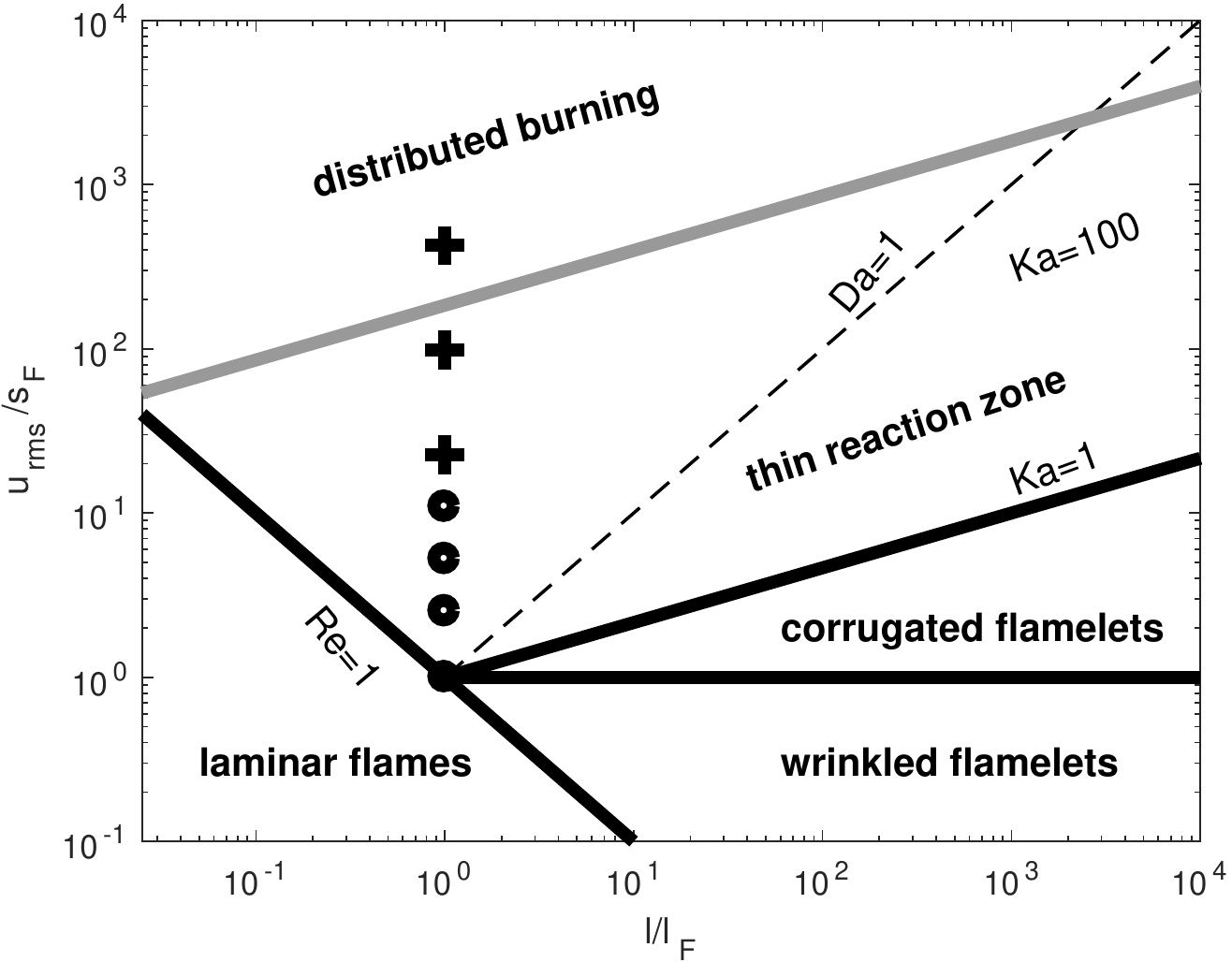}
\caption{Turbulent premixed regime diagram showing the simulations in the present study (plusses) along with our previous simulations (circles) at lower $\Ka$ \citep{AspdenPCI17b}; note that the critical Karlovitz number is shown in grey, and has been placed higher than is usual.}
\label{fig:regime}
\end{figure}

Following our previous studies \citep[e.g.][]{AspdenJFM11,AspdenPCI15,AspdenCnF16},
a canonical periodic-box configuration was used, where
a lean premixed flame was allowed to propagate through 
maintained zero-mean homogeneous isotropic turbulence.
All simulations were run at atmospheric conditions in a high aspect ratio 
domain, with periodic lateral boundary conditions, a free-slip base and outflow at the top. 
Lean premixed hydrogen (equivalence ratio $\varphi=0.4$, Lewis number $\Le\approx0.37$) and
methane ($\varphi=0.7$, $\Le\approx1.0$) were considered.
The freely-propagating hydrogen flame speed and thickness are $s_F=\SI{0.474}{\meter/\second}$ 
and $l_F=\SI{410}{\micro\meter}$;
the laminar methane flame speed and thickness are $s_F=\SI{0.189}{\meter/\second}$ and 
$l_F=\SI{660}{\micro\meter}$.
It was shown in \citet{Aspden08b} that the forcing approach gives approximately 10 integral 
length scales across the domain width.
In all cases the length scale ratio is $\Lambda=l/l_F=1$, consistent with our previous studies 
(e.g.\ \citet{AspdenPCI17b}).  
Three further Karlovitz numbers have been considered $\Ka=108$, 974, and 8767, 
which correspond to velocity ratios $\Upsilon=u^\prime/s_F=22.7$, 98.3, and 425, respectively.
Present simulations and previous simulations are shown respectively as plusses and circles 
  on a regime diagram in figure~\ref{fig:regime}, where the critical Karlovitz number
  for transition to distributed burning has been shown in grey, and placed higher than is usual;
  it should be noted that this critical Karlovitz number is strongly
  dependent on the reactant conditions.

\begin{table}[b!]
\begin{center}
\begin{tabular}{{>{\centering\arraybackslash}p{11mm}|}*{3}{>{\centering\arraybackslash}p{20mm}}|*{3}{>{\centering\arraybackslash}p{20mm}}}
Case & CH$_4$ (108) & CH$_4$ (974) & CH$_4$ (8767) & H$_2$ (108) & H$_2$ (974) & H$_2$ (8767) \\
\hline
$\Upsilon$  &  22.7   &  98.3   &  425  &  22.7   &  98.3   &  425   \\
$\Ka$    &  108   &  974   &  8,767   &  108   &  974   &  8,767  \\
$\Da$    &  0.0440   &  0.0102   &  0.00235  &  0.0440   &  0.0102   &  0.00235 \\
$\Ka_\eta$ &  285   &  2,561   &  23,047   &  350  &  3,149  &  28,339  \\
$\Ka_{\eta_e}$ &  257   &  1,486   &  6,955   &  289   &  1,538  &  7,012  \\
$\Re$ &  157   &  679   &  2,938   &  237   &  1,027   &  4,442  \\
$\Re_e$ &  128   &  229   &  268   &  162   &  245   &  272  
\end{tabular}
\end{center}
\caption{Simulation parameters (based on reactant conditions).}
\label{tab:params}
\end{table}

Most simulations were conducted with a domain size of $10l_F\times10l_F\times80l_F$
discretised on a grid of 192$\times$192$\times$1536 computational cells;
the hydrogen case with the highest $\Ka$ required a larger domain ($10l_F\times10l_F\times120l_F$; 192$\times$192$\times$2304 cells) to accommodate the growth of the flame brush.
This resolution corresponds to 19.2 cells across a thermal thickness, 
which is more than sufficient to resolve these chemical mechanisms \citep{AspdenJFM11,AspdenCnF16}.  
At such high turbulence levels, the Kolmogorov length scale is between 44 and 544 times smaller 
than the thermal thickness of the flame; it is not resolved on this grid
and so these simulations cannot be considered DNS in the sense that the term implies
  well-resolved down to the Kolmogorov length scale.
The effective Kolmogorov length scale (see \citet{Aspden08b} for details)
can be evaluated for this solver, and was found to range between 38 and 67 times smaller than 
the thermal thickness, which indicates that the turbulence that interacts with the flame
(i.e.~at the flame scale) is sufficiently well-resolved (and is maintained by the momentum 
source term through the inertial cascade).
Moreover, as previously argued, turbulence is strongly affected by dilatation,
therefore, close to the flame, the Kolmogorov length is not the value that would be expected from the classical cascade; we argue that scales not represented on the grid have an inconsequential
effect on burning regime and leading-order flame response to turbulence,
and resolving them would be a waste of computational effort.
The simulation parameters are given in table~\ref{tab:params}, where $\Ka_\eta=t_F/t_\eta$
denotes the more conventional Karlovitz number as the ratio of flame time ($t_F=l_F/s_F$)
to Kolmgorov times ($t_\eta=\eta/u_\eta$), $\Re=u^\prime l/\nu$ is the Reynolds number,
and the suffix $e$ denotes effective quantities (i.e.~using $\eta_e$ and $\nu_e$)
evaluated following \cite{Aspden08b}; note that all values are based on reactant conditions.

To establish the consequences of this lack of resolution on the turbulence, simulations of
  non-reacting homogeneous isotropic turbulence have been run at four times the resolution extending
  the length scales at each end of the spectrum separately.  These simulations are presented in
  Appendix~\ref{appA} and demonstrate that the use of maintained turbulence ensures
  the turbulence at the flame scale is the same as that that would have originated from a larger
  integral length scale at the same Karlovitz number (i.e.~has the correct energy dissipation rate)
  despite the limited inertial subrange, and that the scales that are deficient are sufficiently
  smaller than the flame scale that we argue can be considered inconsequential.
  The effective Reynolds number is naturally lower, but the Karlovitz number
  (appropriately defined as above) is unaffected because the energy dissipation rate is
  unaffected by the lack of small-scale resolution.
Note that it is this specific numerical approach (i.e.\ ILES-capable) that means the apparent
  lack of resolution does not invalidate the results; approximately twenty computational cells
  across the flame thermal thickness is sufficient.

An additional simulation was also run to establish the consequences of under-resolving the
  Kolmogorov length scale in the methane case at $\Ka=8767$, which was achieved by adding a
  level of adaptive mesh refinement (AMR) and restarting a calculation from a steady-state
  check point (thereby having nearly 40 computational cells across the laminar thermal thickness);
  the results are presented in Appendix~B.
  The turbulent flame speed was found to be almost unchanged with increased resolution, 
  conditional means of fuel consumption rate and heat release were indistinguishable between
  resolutions, but there were subtle differences between the thickening metric (based on
  temperature gradients).  The higher resolution simulation lends further support to our argument
  that the turbulent scales that are not resolved on the grid are inconsequential as far as
  the flame physics is concerned.

It should be further stressed that these simulations are numerical experiments, and this set of values is not realisable experimentally; in particular, the low Mach number approximation is exploited here to preclude strong compressibility effects and prevent any potential detonation \citep[e.g.][]{PoludnenkoCnF10}. 
Note in particular that the low Mach number approximation is not valid at these conditions;
  these simulations are not intended as a true representation of how these flames would respond, but
  how turbulent flames at the same Karlovitz and Damk\"ohler numbers would respond if conditions
  could be constructed at low Mach number.  Furthermore, viscous heating has not been
  included, which would have the potential to lead to a significant rise in reactant temperature
  as a result of continual energy injection by the momentum forcing term.
Despite the limitations of this configuration, especially at the highest $\Ka$, these simulations capture turbulence-flame interactions during the transition towards the limiting case of flame propagation driven by turbulent mixing, i.e.~distributed burning, and are of significant interest and relevance to the transition away from the thin reaction zone with increasing levels of turbulence.  Possible steps to realise distributed burning experimentally are discussed in section~\ref{sec:conclusions}.

\section{Results}

\subsection{Flame response overview}

\begin{figure}[ht!]
\centering
\iflowres
\includegraphics[width=0.48\textwidth]{images/ch4-ka108-lr}
\includegraphics[width=0.48\textwidth]{images/h2-ka108-lr}\\
\includegraphics[width=0.48\textwidth]{images/ch4-ka974-lr}
\includegraphics[width=0.48\textwidth]{images/h2-ka974-lr}\\
\includegraphics[width=0.48\textwidth]{images/ch4-ka8767-lr}
\includegraphics[width=0.48\textwidth]{images/h2-ka8767-lr}\\
\else
\includegraphics[width=0.48\textwidth]{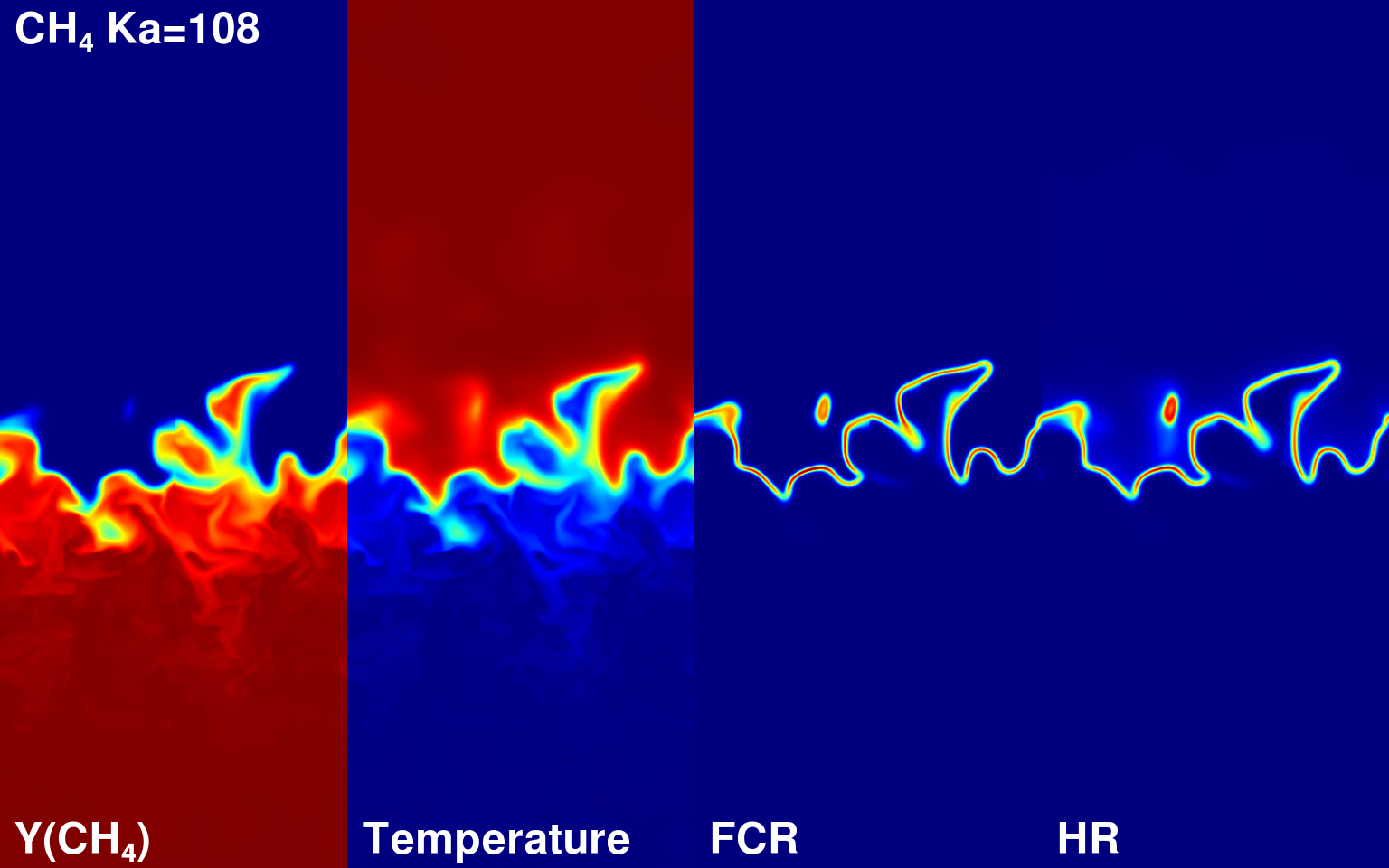}
\includegraphics[width=0.48\textwidth]{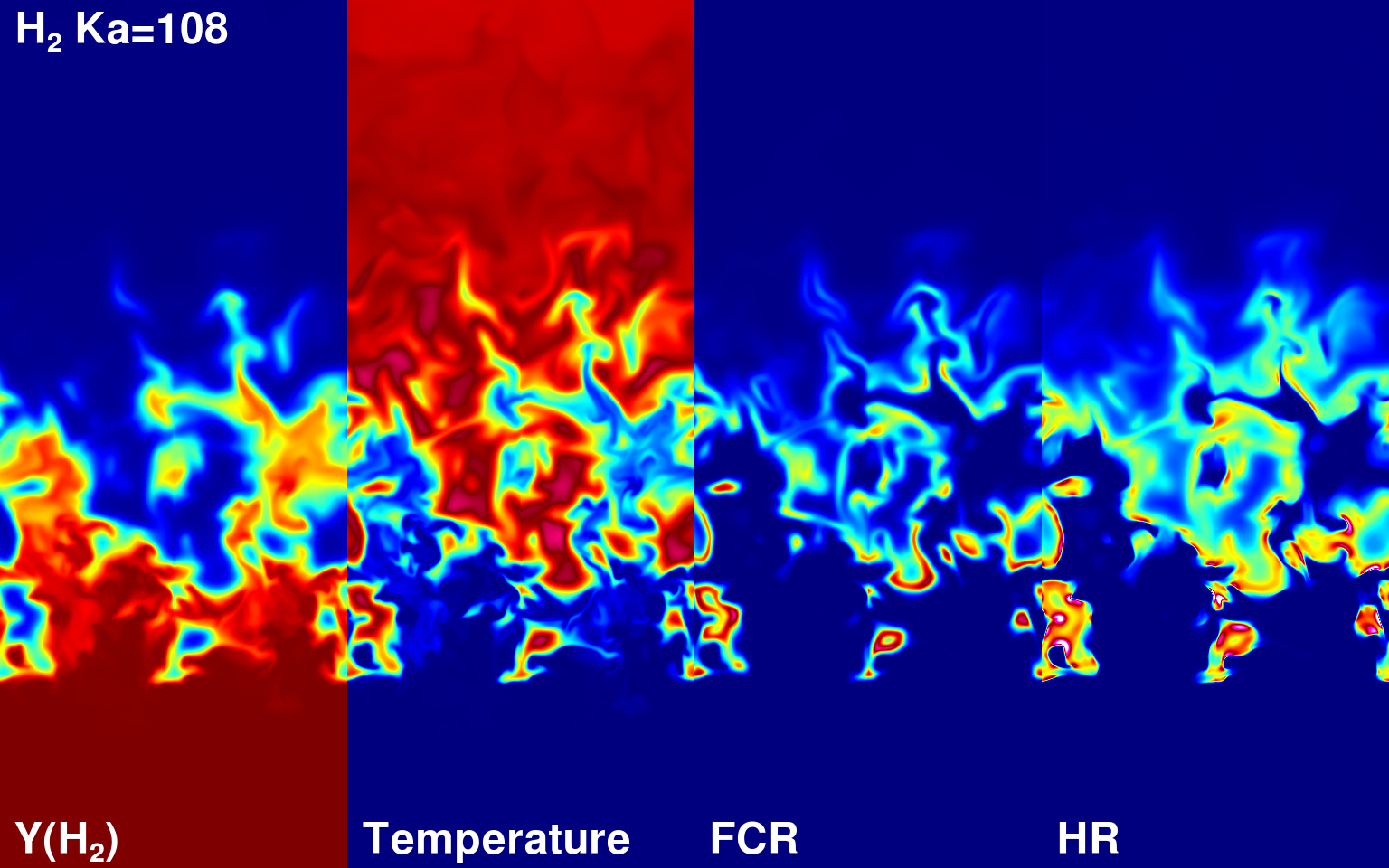}\\
\includegraphics[width=0.48\textwidth]{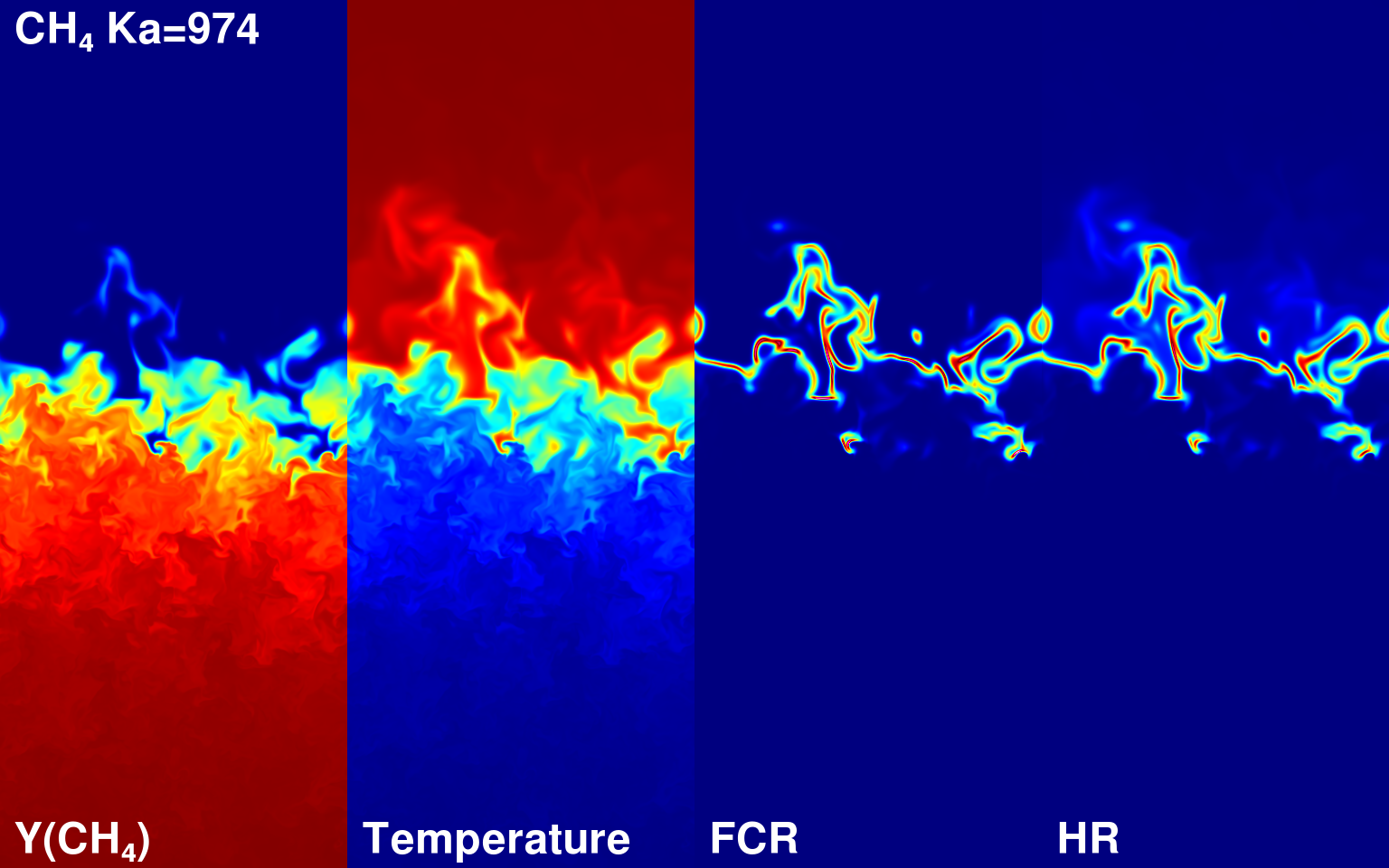}
\includegraphics[width=0.48\textwidth]{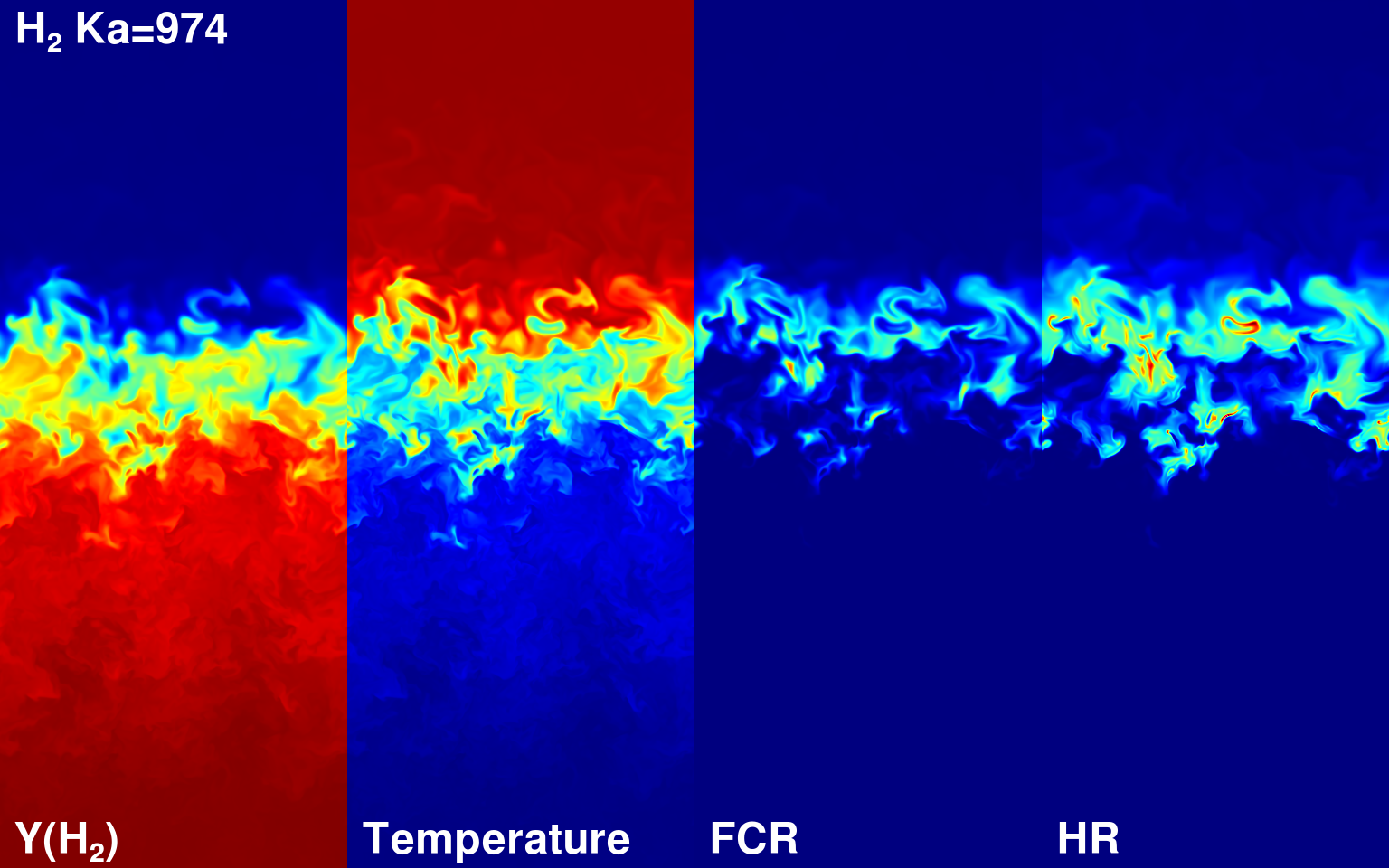}\\
\includegraphics[width=0.48\textwidth]{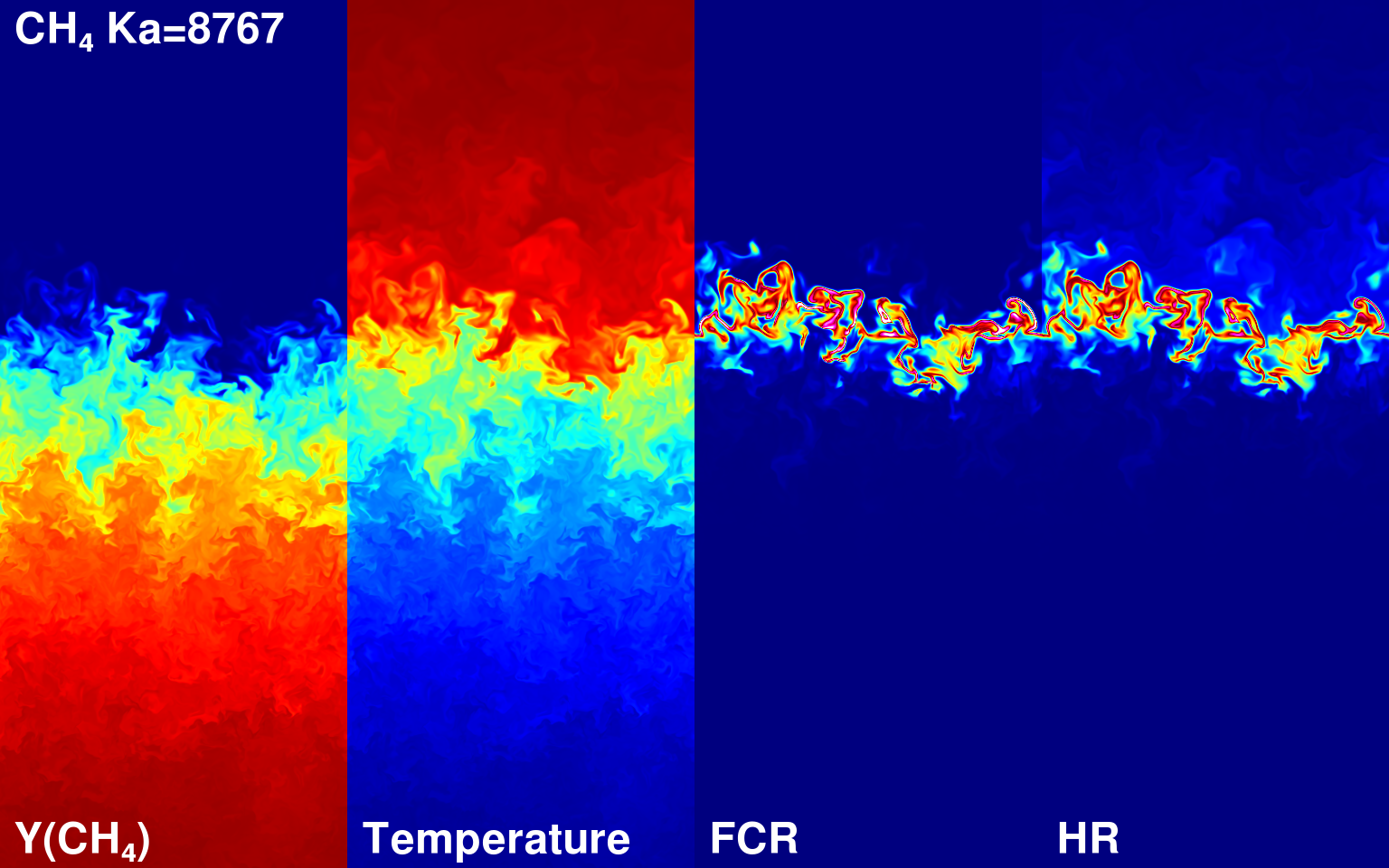}
\includegraphics[width=0.48\textwidth]{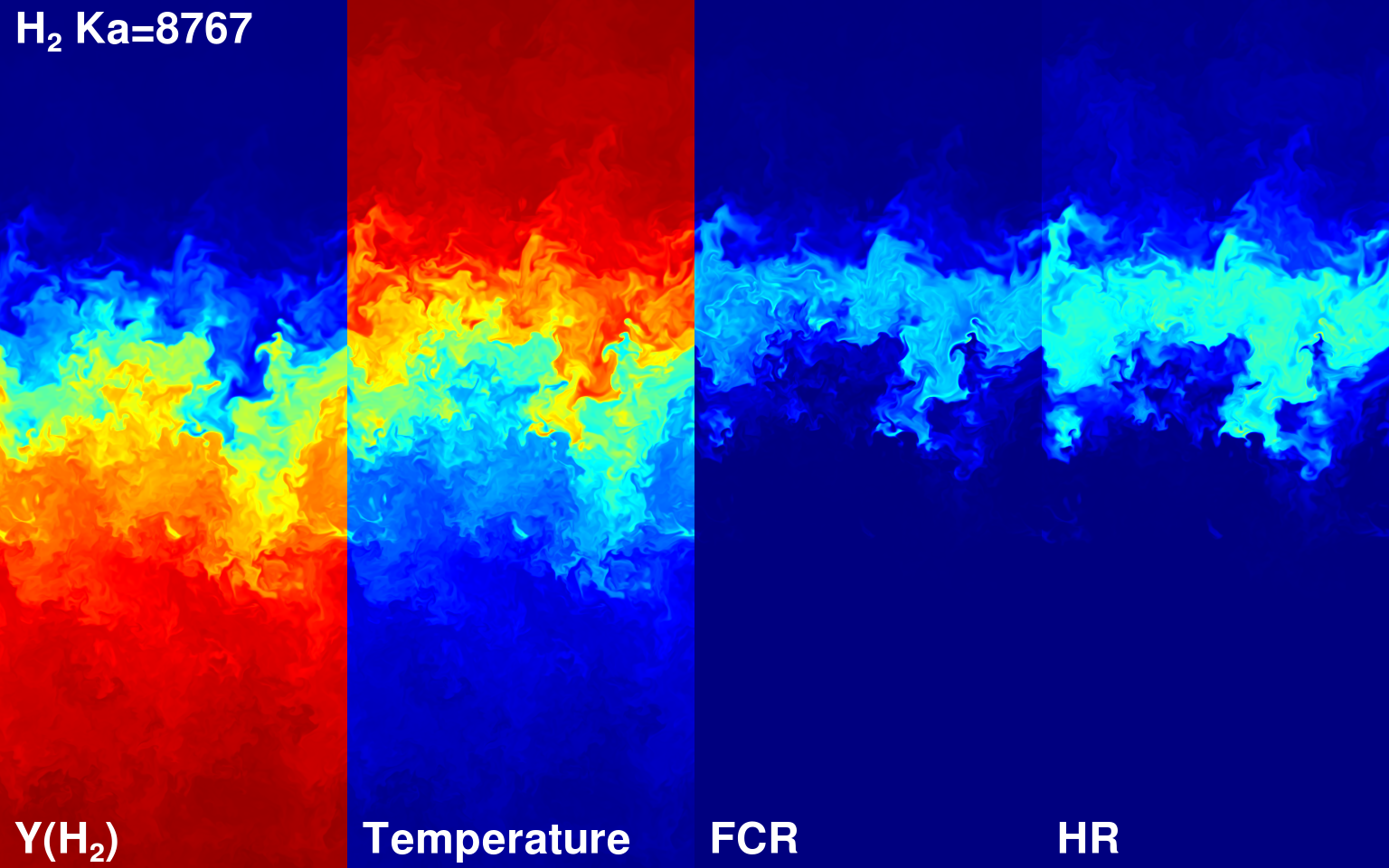}\\
\fi
\includegraphics[width=0.48\textwidth]{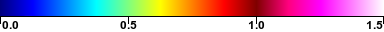}
\caption{Slices of fuel mass fraction, temperature, fuel consumption rate, and heat release for CH$_4$ and H$_2$ flames at $\Ka=108$, 974 and 8767, respectively.  Each panel of each image shows $20l_F\times50l_F$ (note that periodicity has been exploited to stitch together $x$-$z$ and $y$-$z$ planes to show more flame surface).}
\label{fig:slices}
\end{figure}

A general overview of flame response to high $\Ka$ turbulence is presented in
figure~\ref{fig:slices}, which depicts slices of fuel mass fraction, temperature,
fuel consumption rate (FCR), and heat release rate (HR), normalised by the
corresponding laminar values (note for for hydrogen, the FCR and HR are normalised
by ten times the laminar value to allow for the enhanced reaction rates due to the
thermodiffusive instability).

Methane at $\Ka=108$ appears to be similar to moderate $\Ka$ \citep[see the $\Ka=36$ case in][]{AspdenCnF16}; the flame surface is convoluted but smooth, generally similar to the laminar flame, with a decrease in reaction rates correlated with high positive curvature, which was attributed to atomic hydrogen diffusion \citep{Echekki96,AspdenCnF16}. The temperature field shows clear evidence of turbulent mixing, but restricted to the preheat region.

Methane at $\Ka=974$ continues the trend of increased turbulent mixing in the preheat region, but the reaction rates appear to be changing; while not immediately apparent in the image, the peak value exceeds the laminar flame values in places, and there appears to be greater variability along the flame surface, which is becoming more convoluted -- again, this can be attributed to atomic hydrogen diffusion \citep{Echekki96,AspdenCnF16}.

\begin{figure}
\centering
\iflowres
\includegraphics[height=44mm]{images/ch4-ka8767-lam-lr}
\includegraphics[height=44mm]{images/h2-ka8767-lam-lr}
\includegraphics[height=44mm]{images/sneFCR-lr}
\else
\includegraphics[height=44mm]{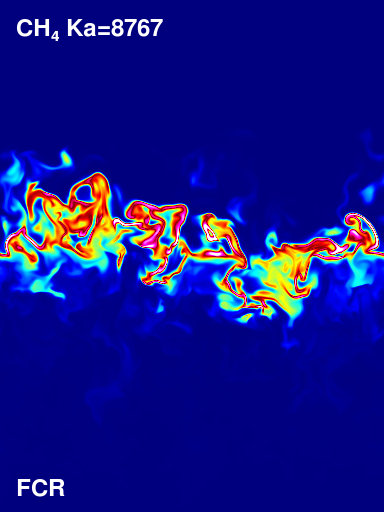}
\includegraphics[height=44mm]{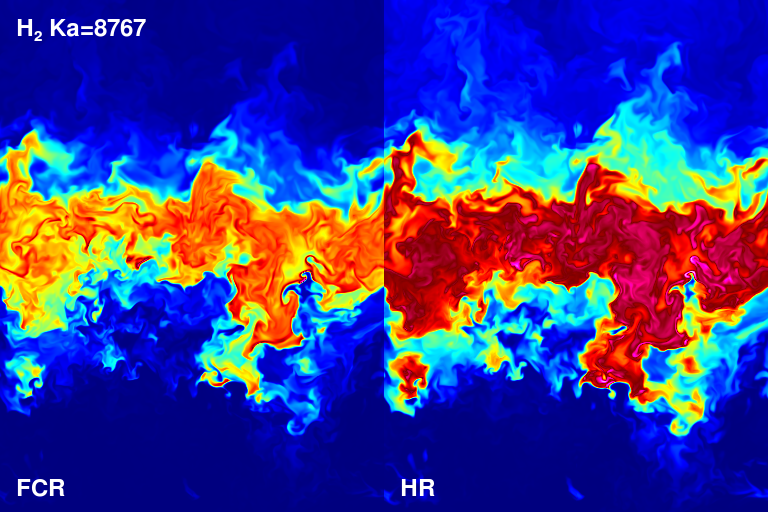}
\includegraphics[height=44mm]{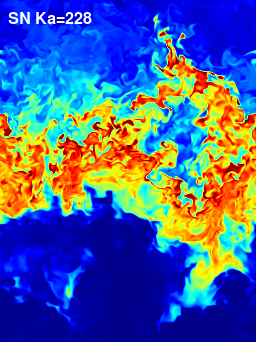}
\fi
\includegraphics[width=0.48\textwidth]{images/colourBar}
\caption{Slices of FCR for CH$_4$, and FCR and HR for H$_2$ at $\Ka=8767$ (each panel shows approximately $20l_F\times30l_F$); for H$_2$ the normalisation is four times the laminar value (rather than ten).  The reaction rate from the distributed supernova flame \citep{Aspden08a} is shown for comparison (the panel shows approximately $40l_F\times60l_F$), which is normalised by one-fifth of the corresponding laminar value.}
\label{fig:slices-lam}
\end{figure}

Methane at $\Ka=8767$ shows different behaviour than the lower $\Ka$ cases; in addition to the mixing observed in the preheat region, there are indications in the temperature field that there is an onset of turbulent mixing in the post-flame region.  Interestingly, the reaction rates appear slightly increased in places (see magenta/white regions; an enlarged image is shown in figure~\ref{fig:slices-lam}), but are not broadened by turbulence.

Hydrogen at $\Ka=108$ presents significant thermodiffusively-unstable behaviour \citep[e.g.][]{Trouve94,Baum94}, and continues the trend from that observed in \citet{AspdenJFM11,AspdenPCI15}; turbulence exaggerates the thermodiffusive instability, creating small-scale structures with higher curvature than at lower $\Ka$, resulting in more intense burning over a broad flame brush (localised heat release rates in excess of fifteen times the laminar value are observed; shown by the white regions).  The decorrelation between fuel consumption and heat release rates (reported by \citet{ChenPCI00} and \cite{AspdenPCI15}, and demonstrated to be due to atomic hydrogen diffusion \citet{AspdenPCI17a}) is present, further indicating persistence of preferential diffusion at this $\Ka$.  Super-adiabatic temperatures still exist in the near post flame region.  The temperature field presents limited turbulent mixing; the thermodiffusive instability leads to a resistance to turbulent mixing.  

At $\Ka=974$, the hydrogen flame presents the first evidence of a change in behaviour; the thermodiffusively unstable structures at lower $\Ka$ are no longer observed, reaction rates have generally decreased, and the temperature field shows significant evidence of turbulent mixing in the preheat region.

At the highest $\Ka$, the hydrogen flame presents significantly different behaviour to all of the other cases; there is now substantial turbulent mixing throughout the flame, there is no evidence of the thermodiffusive instability, the reaction rates are substantially lower than the other two hydrogen cases, and distributed relatively smoothly across a region that is over ten thermal thicknesses across.  This behaviour is consistent with the distributed supernova flame presented in \citet{Aspden08a}; figure~\ref{fig:slices-lam} reinforces this similarity by comparing the reaction rates from the distributed supernova flame with hydrogen fuel consumption rate and heat release
(now normalised by four times the laminar values rather than ten).
The peak reaction rates in the hydrogen case are lower than those of the moderate turbulent cases,
but remain higher than in the one-dimensional laminar flame.
This is in contrast to the supernova flame that showed a decrease in local fuel consumption
  rate; this can be explained as the suppression of Lewis number effects, 
  and will be discussed further in section~\ref{sec:conclusions}.

\subsection{Turbulent flame speeds}

\begin{figure}[th!]
\centering
\includegraphics[width=0.48\textwidth]{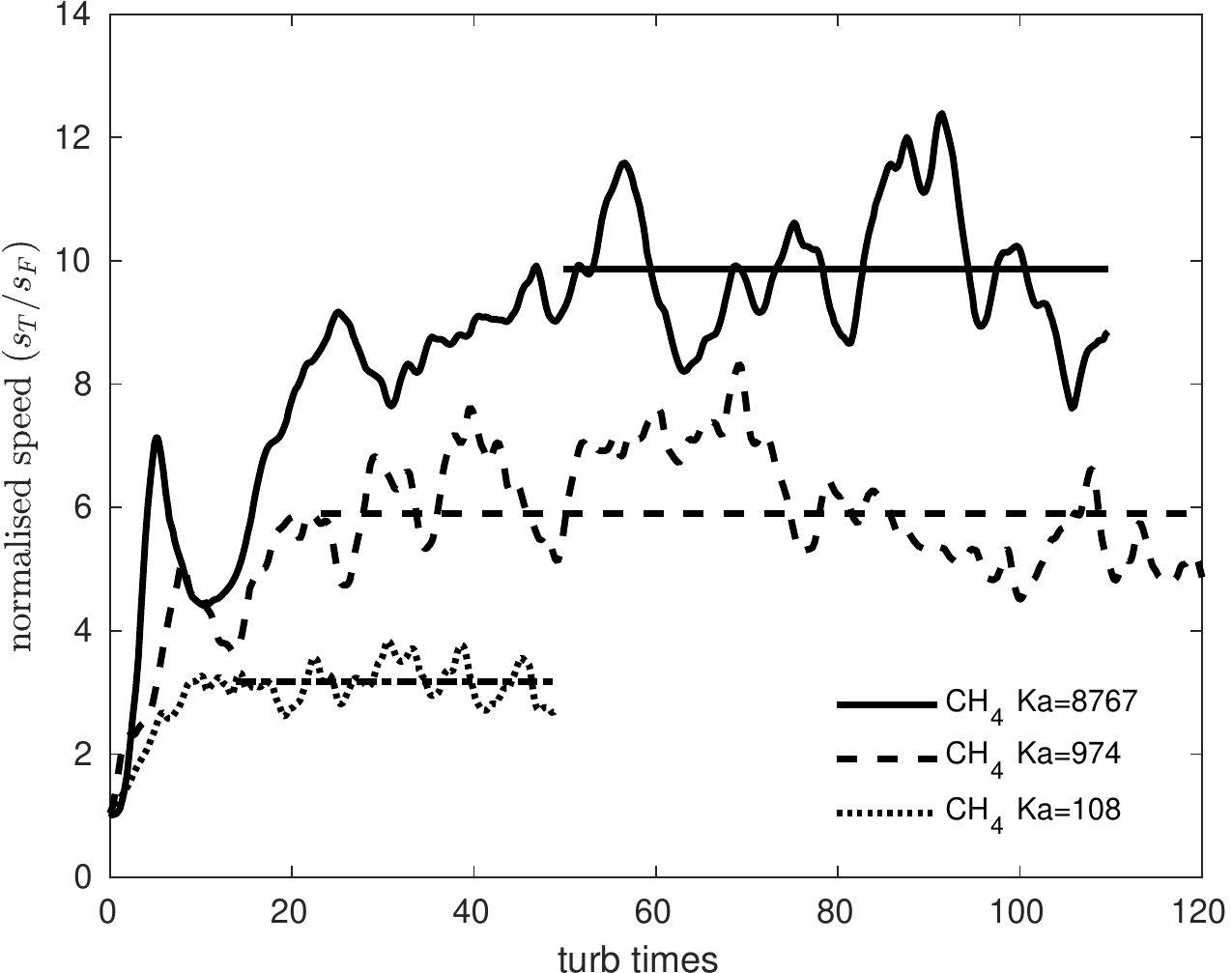}
\includegraphics[width=0.48\textwidth]{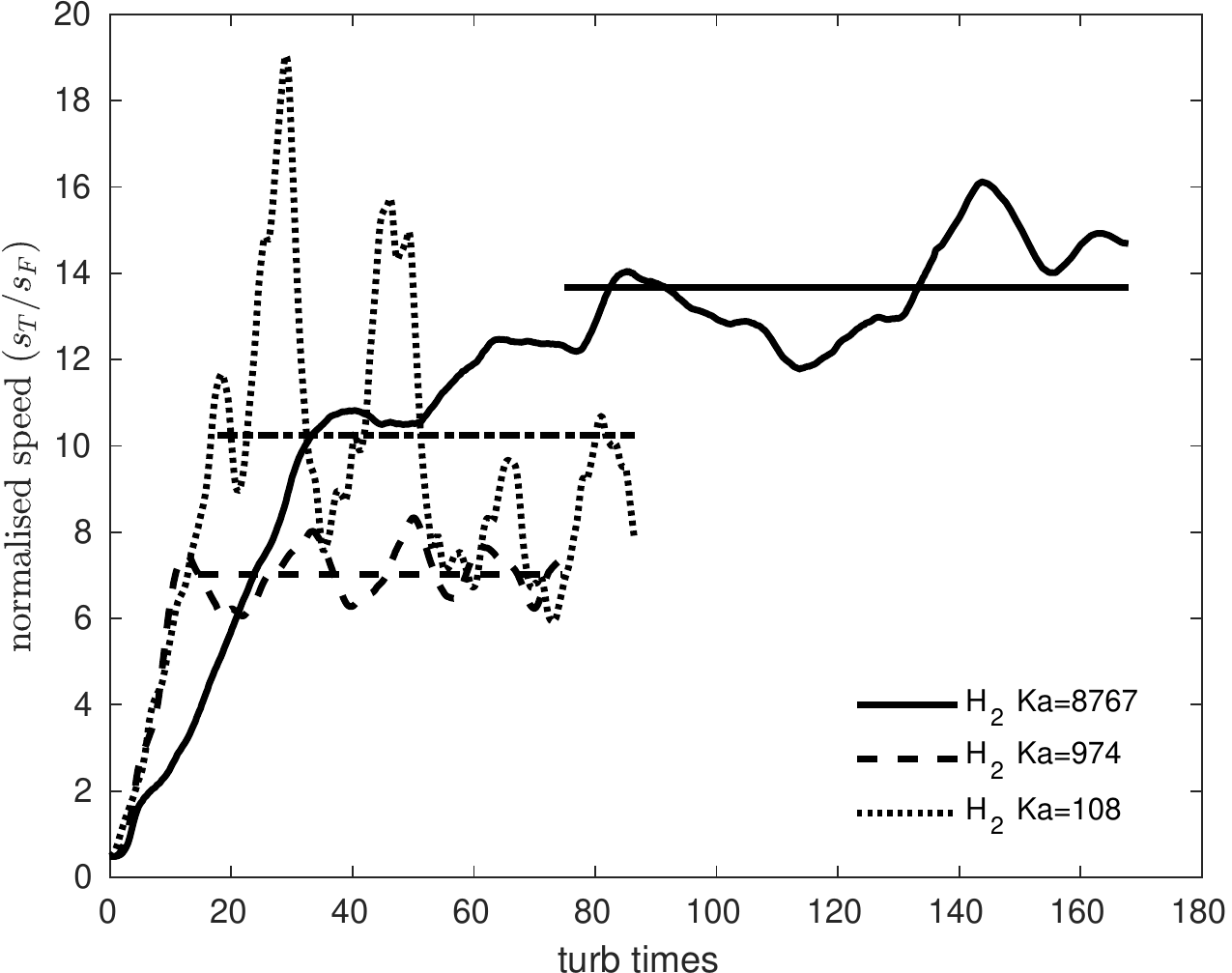}\vspace{2mm}
\includegraphics[width=0.48\textwidth]{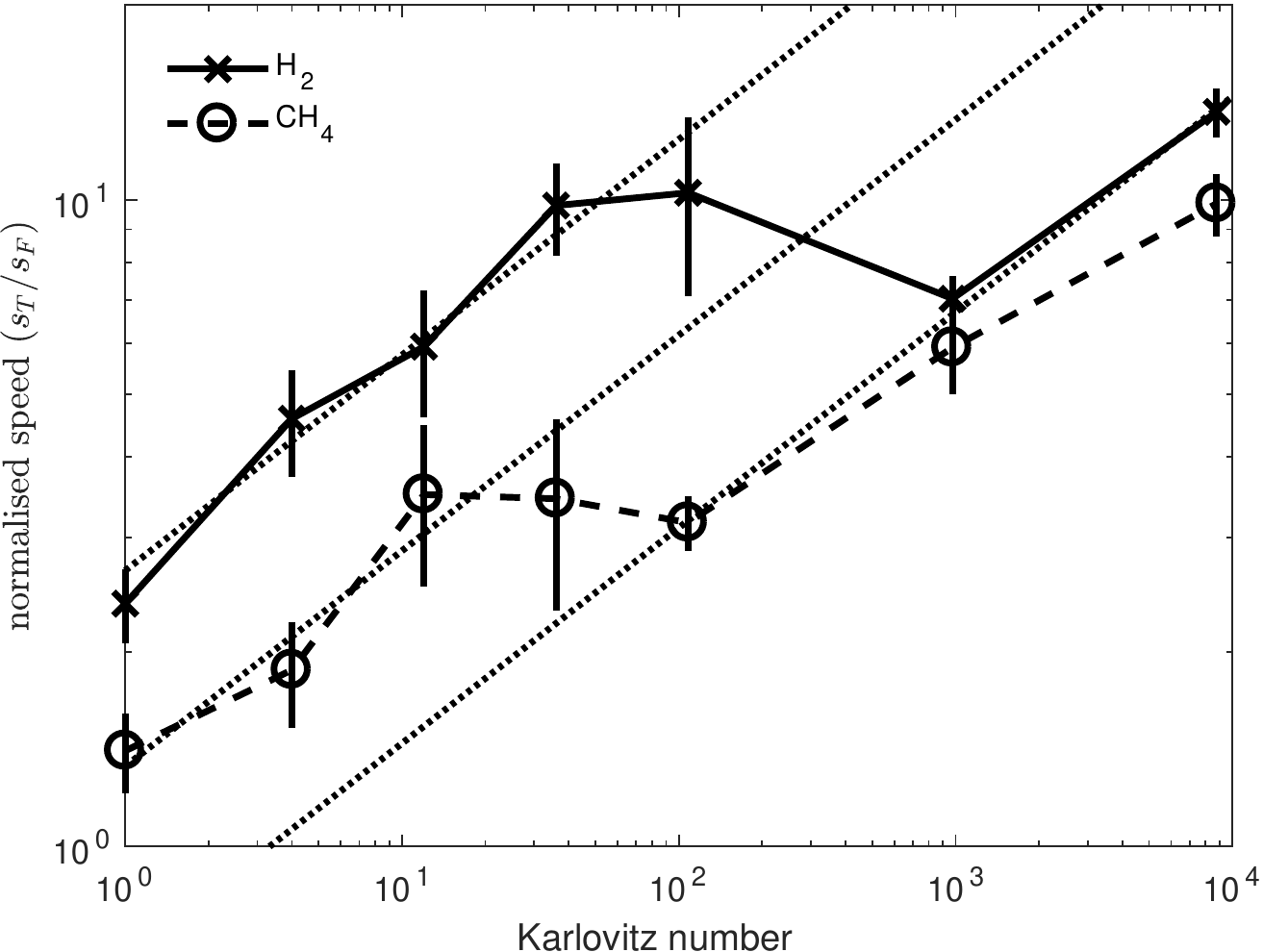}
\caption{Top: normalised turbulent flame speeds $s_T/s_F$ as a function of turbulent eddy turnover times for methane (left) and hydrogen (right); horizontal lines denote the average and period treated as steady-state.  Bottom: Normalised turbulent flame speed as a function of $\Ka$; the vertical lines show one standard deviation about the mean, and the dotted lines show the scaling $s_T/s_F\propto\Ka^{1/3}$ (for fixed $\Lambda$), with three different constants of proportionality.}
\label{fig:st}
\end{figure}

Turbulent flame speeds are shown both as a function of time and
as a function of $\Ka$ in figure~\ref{fig:st} (vertical lines in the latter
denote the standard deviation after reaching a statistically-steady state). 
Note that at such extreme levels
of turbulence, tens of integral length eddy turnover times are required for the flame to reach
a statistically-steady state, and closer to 100 for the highest cases.
The horizontal lines denote the mean and averaging period; in the following sections,
all data has been temporally averaged in time using around 100 time points evenly distributed
over the corresponding time periods.
Power law scaling (as a function of $\Ka$) can be
derived in analogy with \citet{Damkohler40} for the distributed limit of turbulence-driven 
mixing, which is shown by the dotted lines
(see also \citet{Peters00}).
Predicting a turbulent flame speed $s_T=\sqrt{D_T/\tau_T}$, where $D_T=\alpha u^\prime l$, is
a turbulent diffusion for some constant $\alpha$ and (turbulent) chemical time scale $\tau_T$ 
(both to be determined), three equivalent expressions can be derived
\begin{equation}
  \frac{s_T}{s_F}\propto\Lambda^{2/3}\,\Ka^{1/3},
  \qquad
  \frac{s_T}{s_F}\propto\Lambda\,\Da^{-1/2},
  \qquad\text{or}\qquad
  \frac{s_T}{s_F}\propto\Lambda^{1/2}\,\Upsilon^{1/2},
\end{equation}
where the constant of proportionality in all three cases is $\alpha^{1/2}\Da_T^{-1/2}$;
note the latter can be written as $(\Re/\Re_F)^{1/2}$ for $\Re_F=s_Fl_F/\nu$.
It is as yet unclear why both hydrogen and methane flames appear to follow this scaling law
in a regime where it is not intended to apply (other than a simple dimensional necessity)
followed by an apparent transition, after which the flame speeds may again follow the scaling
law, especially since distributed burning behaviour is only observed for hydrogen at $\Ka=8767$.
It should be be borne in mind that the present measure of turbulent flame speed is a
  global-consumption-based metric, and will depend on the size of the domain (which is usually
  not considered). Extending the domain
  without changing the integral length scale is unlikely to result in a simple periodic
  replica of the present flame; there is a greater volume into which the flame can develop,
  leading to greater flame surface area, and therefore greater turbulent flame speed
  by the present global-consumption-based metric.

\subsection{Thickening factor}

\begin{figure}
\centering
\makebox[0.5\textwidth][c]{Methane flames}\makebox[0.5\textwidth][c]{Hydrogen flames}
\includegraphics[width=0.48\textwidth]{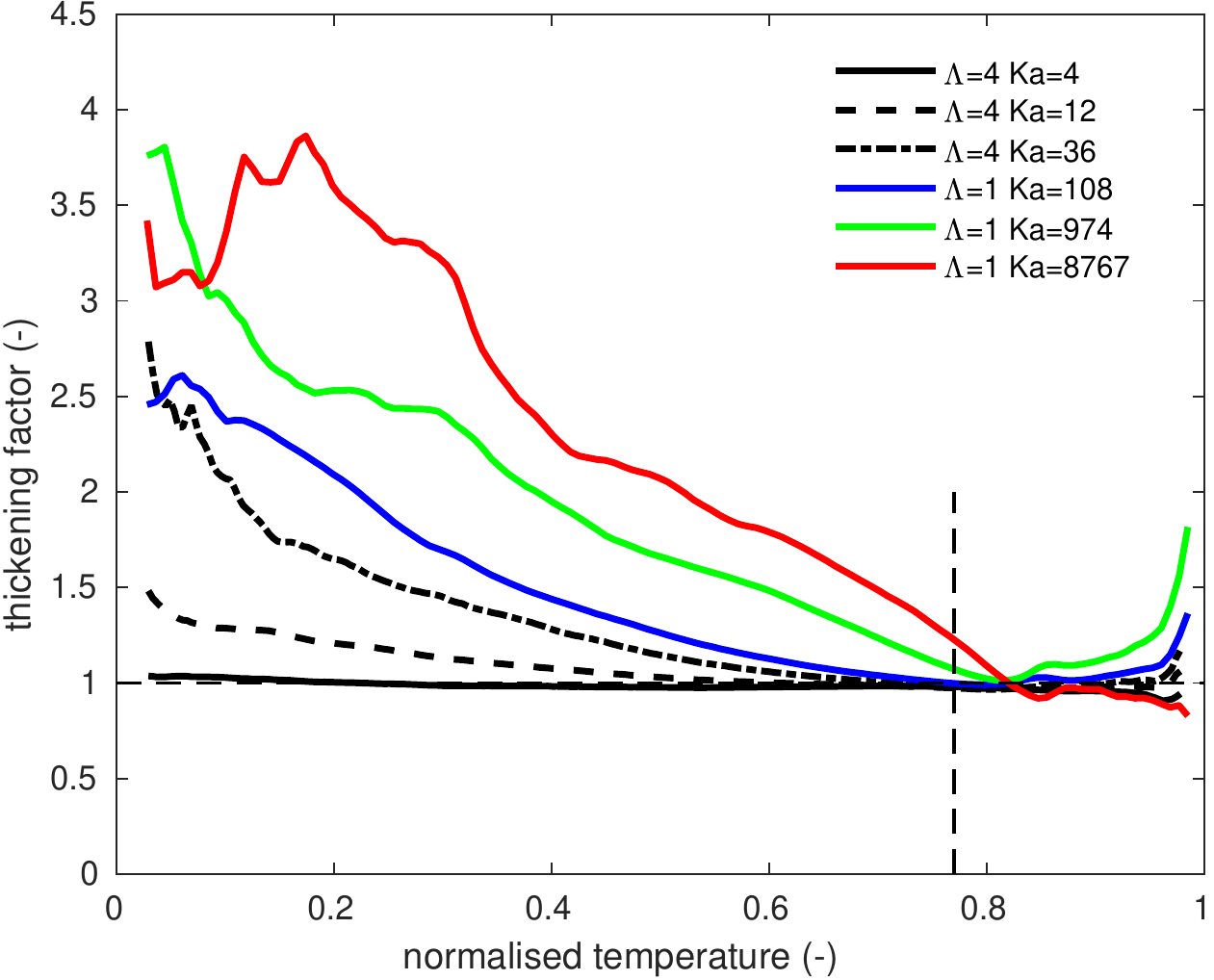}
\includegraphics[width=0.48\textwidth]{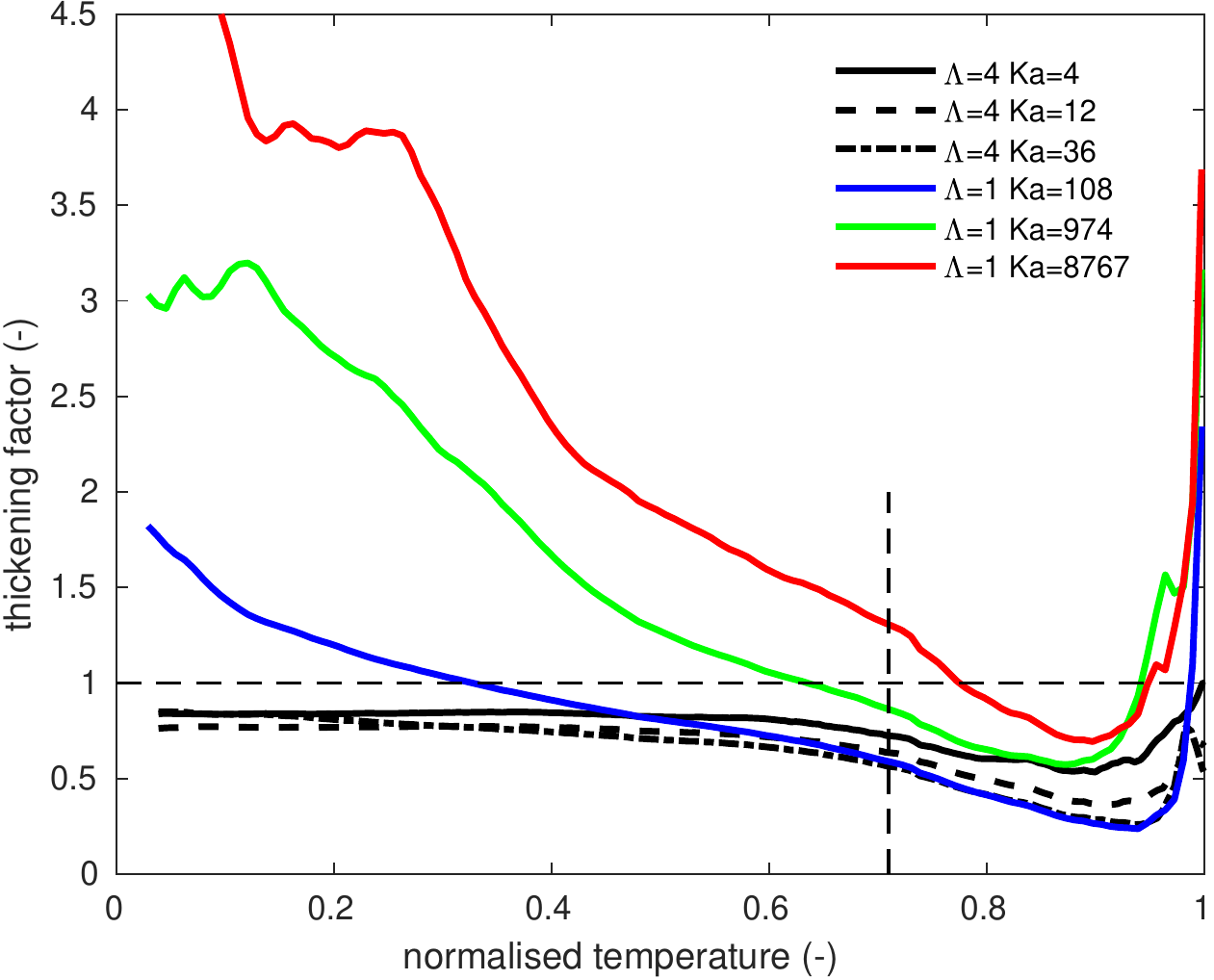}
\caption{Thickening factor $\Theta(T)$ for CH$_4$ (left) and H$_2$ (right) compared with
previous lower $\Ka$ cases from \cite{AspdenPCI15,AspdenCnF16}, which are shown in black.}
\label{fig:theta}
\end{figure}

A local thickening factor was previously defined \citep{AspdenCnF16} analogously to thermal
thickness as the ratio of the conditional means of temperature gradients as
\begin{equation}
\Theta_n(T)=\frac{<\nabla T(\xi)\,|\,\xi=T>_{\ssKa=1}}{<\nabla T(\xi)\,|\,\xi=T>_{\ssKa=n}},
\end{equation}
where the normalisation by the conditional mean at $\Ka=1$ is used (in preference to the laminar
profile) to account for the thermodiffusive instability in the hydrogen flames.  In that paper,
the methane flames were shown to be broadened in the preheat region,
unlike the hydrogen flames, which became progressively thinner with increasing $\Ka$ 
due to generation of more highly-curved flame surface by turbulence thereby enhancing the
thermodiffusive effects.  The thickening factor in the present flames are compared in
figure~\ref{fig:theta} with these previous lower $\Ka$ cases; 
the methane flames are broadened further in the preheat region
with increasing $\Ka$, but seem to remain thin at $\Ka=8767$ in the post-flame region 
(the vertical dashed line indicates the location of peak HR). The thinning trend in hydrogen is
reversed in the preheat region at higher $\Ka$, but remain thinner in the post-flame region.  This 
thinning in the post-flame region may be due in part to the normalisation, where the 
temperature profile $\Ka=1$ flame (as for the laminar flame) has a long tail at high
temperatures (i.e.~small gradients), resulting in a smaller $\Theta$ due a modified post-flame
structure at high $\Ka$.
Furthermore, it appears that the thickening factor does not reflect the visual thickening 
observed in figure~\ref{fig:slices}; as the turbulence intensity increases, the flame becomes less 
flamelet-like, turbulent mixing maintains strong local gradients distributed over a broad
region in space, which is compounded by conditioning on temperature without accounting for the
broadening of the temperature field itself.
At the highest $\Ka$, this metric becomes less reliable in the preheat region due to resolution
limitations; the simulation with refined computational grid demonstrates that this sensitivity
is predominantly observed in the preheat region, but is slight in the region of the flame
(see figure~\ref{fig:convCH4} in the appendix).

\subsection{Reaction rates}

\begin{figure}[th!]
\centering
\makebox[0.5\textwidth][c]{Methane flames}\makebox[0.5\textwidth][c]{Hydrogen flames}
\includegraphics[width=0.48\textwidth]{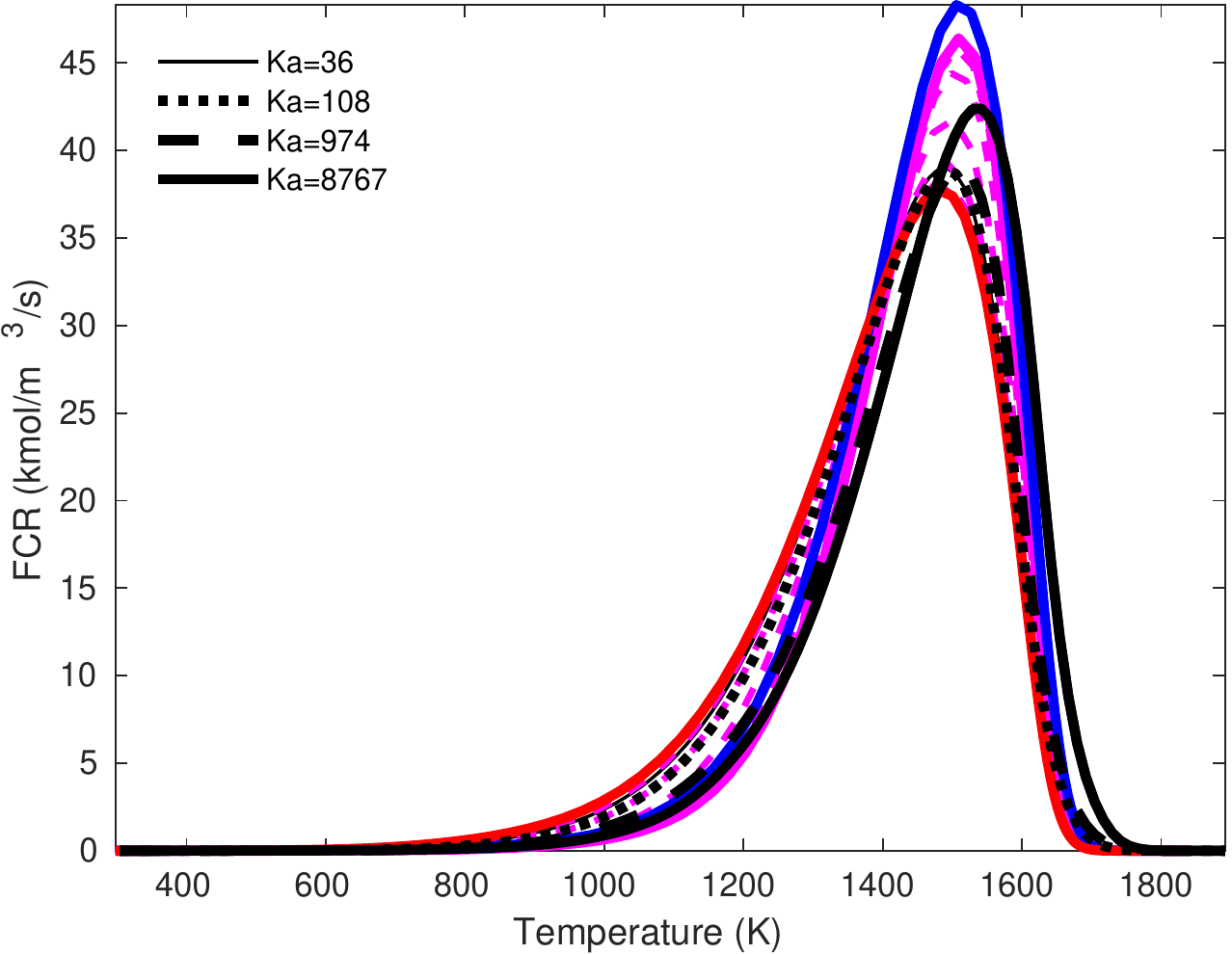}
\includegraphics[width=0.48\textwidth]{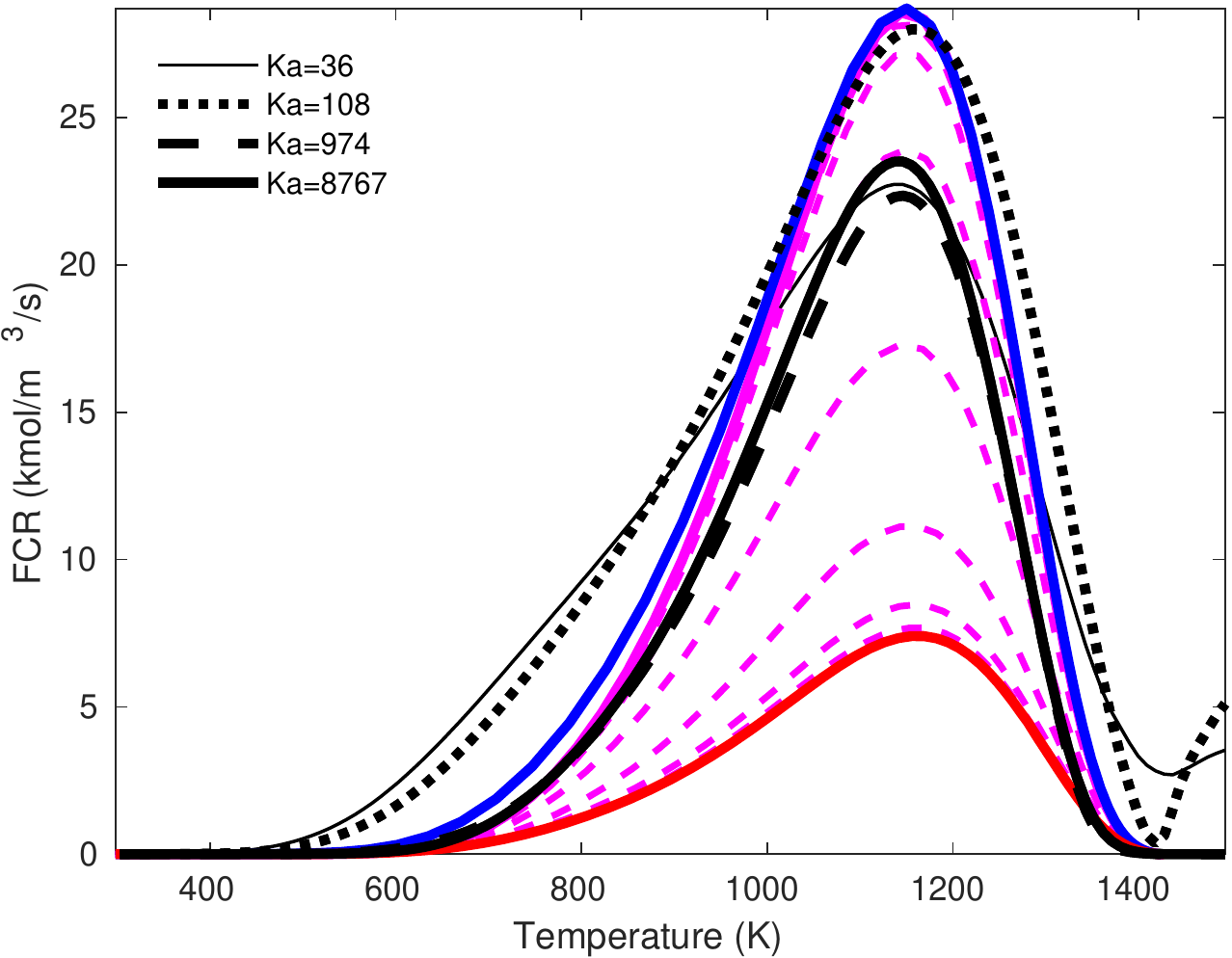}
\includegraphics[width=0.48\textwidth]{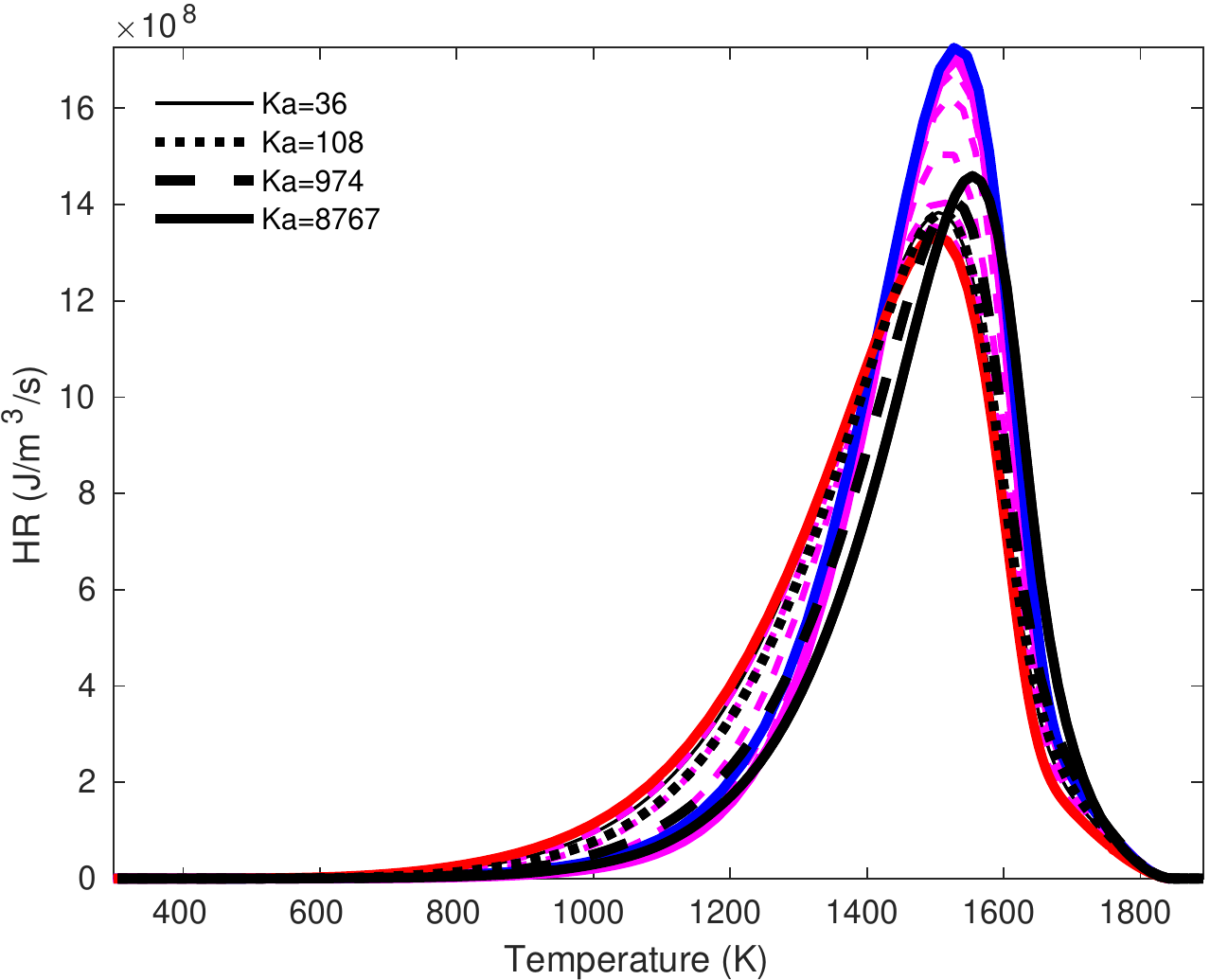}
\includegraphics[width=0.48\textwidth]{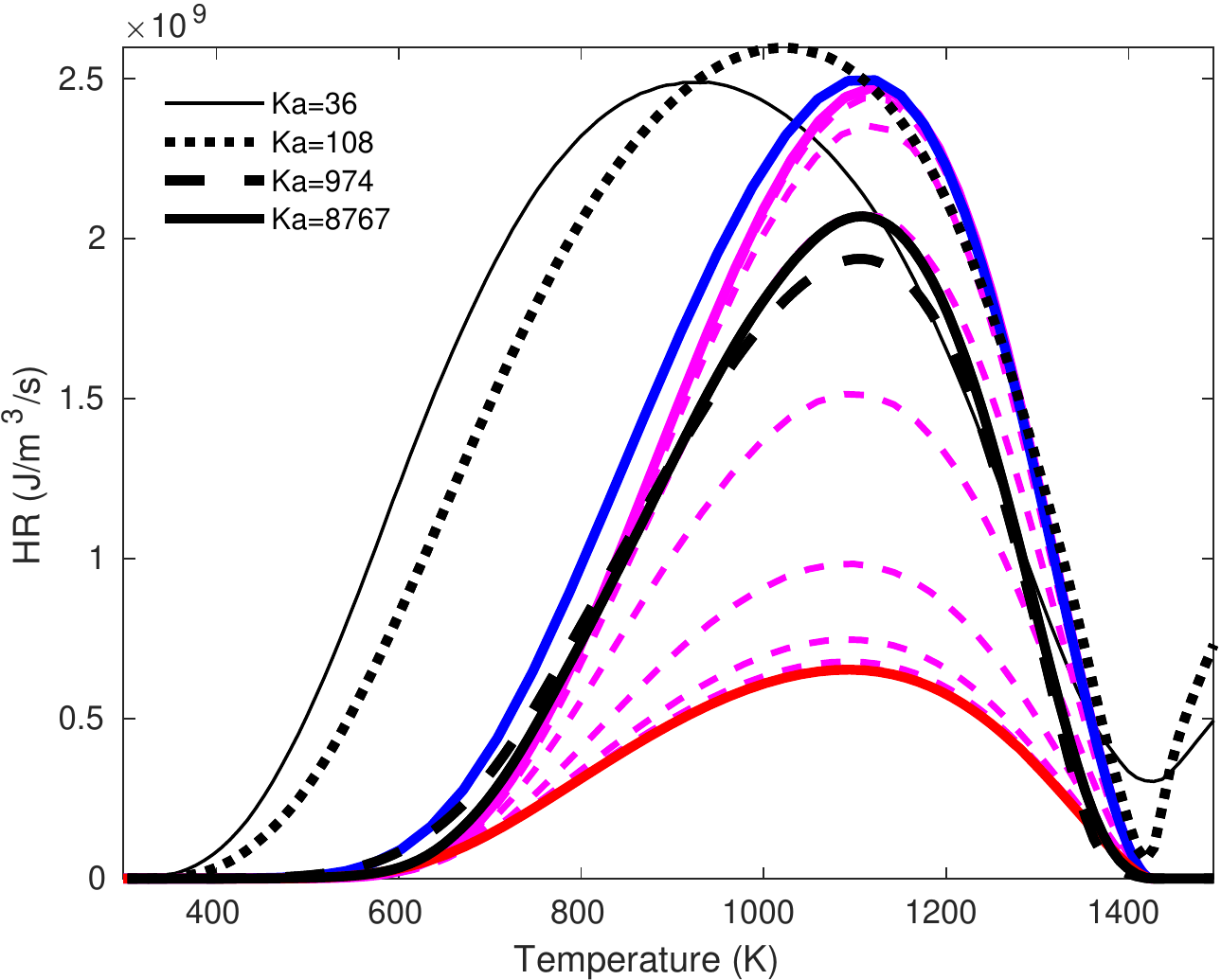}
\caption{Conditional means of fuel consumption rate and heat release for CH$_4$ (left) and H$_2$ (right) flames.  
Black lines are the conditional means from the simulations, red is the one-dimensional laminar 
flame profile, blue is the one-dimensional unity Lewis number profile, and the dashed magenta lines 
are one-dimensional laminar flames with an added turbulent diffusion of increasing magnitude with
the limiting case shown as a solid line.}
\label{fig:cm-fcr-hr}
\end{figure}

The response of reaction rates to turbulence is considered using 
conditional means of fuel consumption rate and heat release rate in
figure~\ref{fig:cm-fcr-hr}.
One-dimensional flame solutions are also presented for comparison:
the red line denotes the laminar flame and the blue line denotes the unity Lewis number flame.
The dashed magenta lines were obtained by one-dimensional calculations of the laminar flame
where each diffusion coefficient has been supplemented by a constant turbulent diffusion
(i.e.~the diffusion coefficient $D_i$ for species $i$ was increased by $D_T$, where $D_T$ is 
a turbulent diffusion coefficient, with a corresponding thermal term to ensure unity Lewis number of the supplementary terms); $D_T$ was gradually increased in magnitude
(giving the different dashed magenta lines) until the profile stopped changing, and that 
limiting case is shown by the solid magenta line.

At low-to-moderate $\Ka$, the methane profile is close to the laminar profile, but at the highest
$\Ka$ the profile appears to have shifted toward higher temperatures, with a slight increase in
peak magnitude (consistent with figure~\ref{fig:slices}).
A more pronounced response is observed in the hydrogen flames; at low-to-moderate $\Ka$, there is substantial heat release rate in the preheat region (due to the thermodiffusive instability
exaggerated by moderate turbulence), whereas at higher $\Ka$ the distribution is narrower,
with a peak that again has shifted to higher temperatures, consistent with the peak temperature
in the one-dimensional profiles, again indicative of suppression of Lewis number effects.
Importantly, note the similarity between the hydrogen flames at $\Ka=974$ and 8767; 
this suggests that the conditional mean is tending to a limiting distribution, which we argue
is to be anticipated as a characteristic of distributed burning.

\subsection{Species mass fraction distributions}

\begin{figure}
\centering
\makebox[0.5\textwidth][c]{Methane flames}\makebox[0.5\textwidth][c]{Hydrogen flames}
\includegraphics[width=0.48\textwidth]{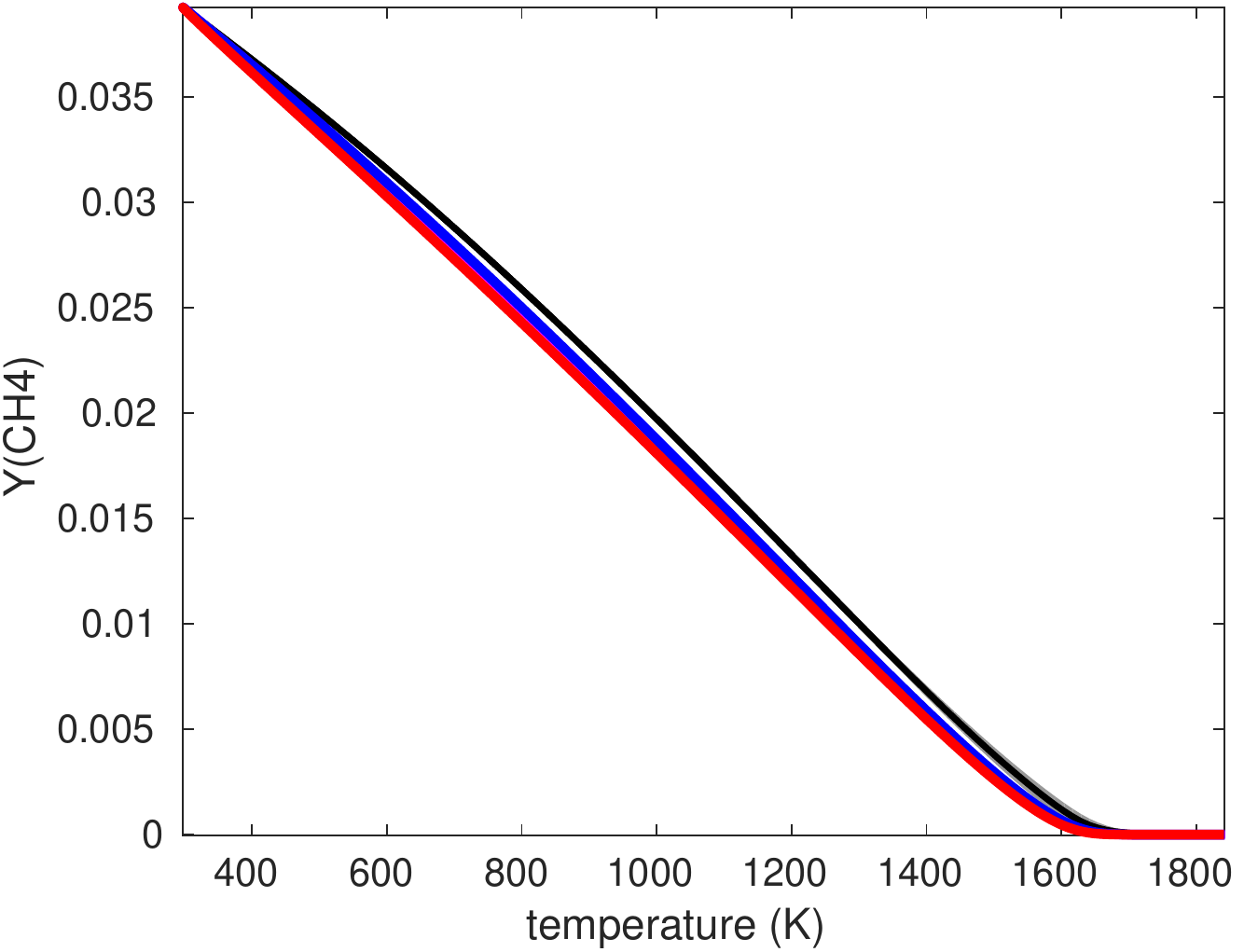}
\includegraphics[width=0.48\textwidth]{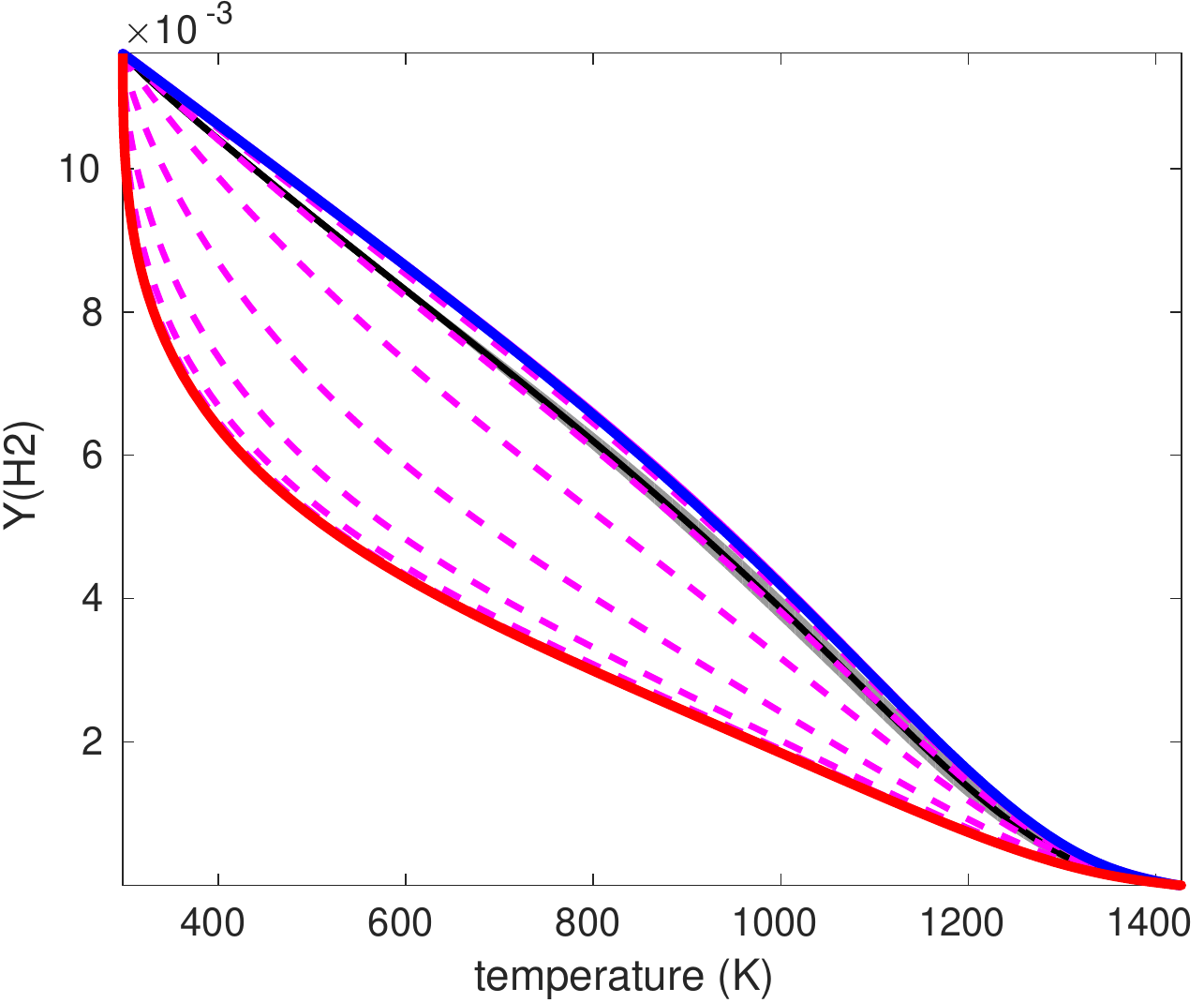}
\includegraphics[width=0.48\textwidth]{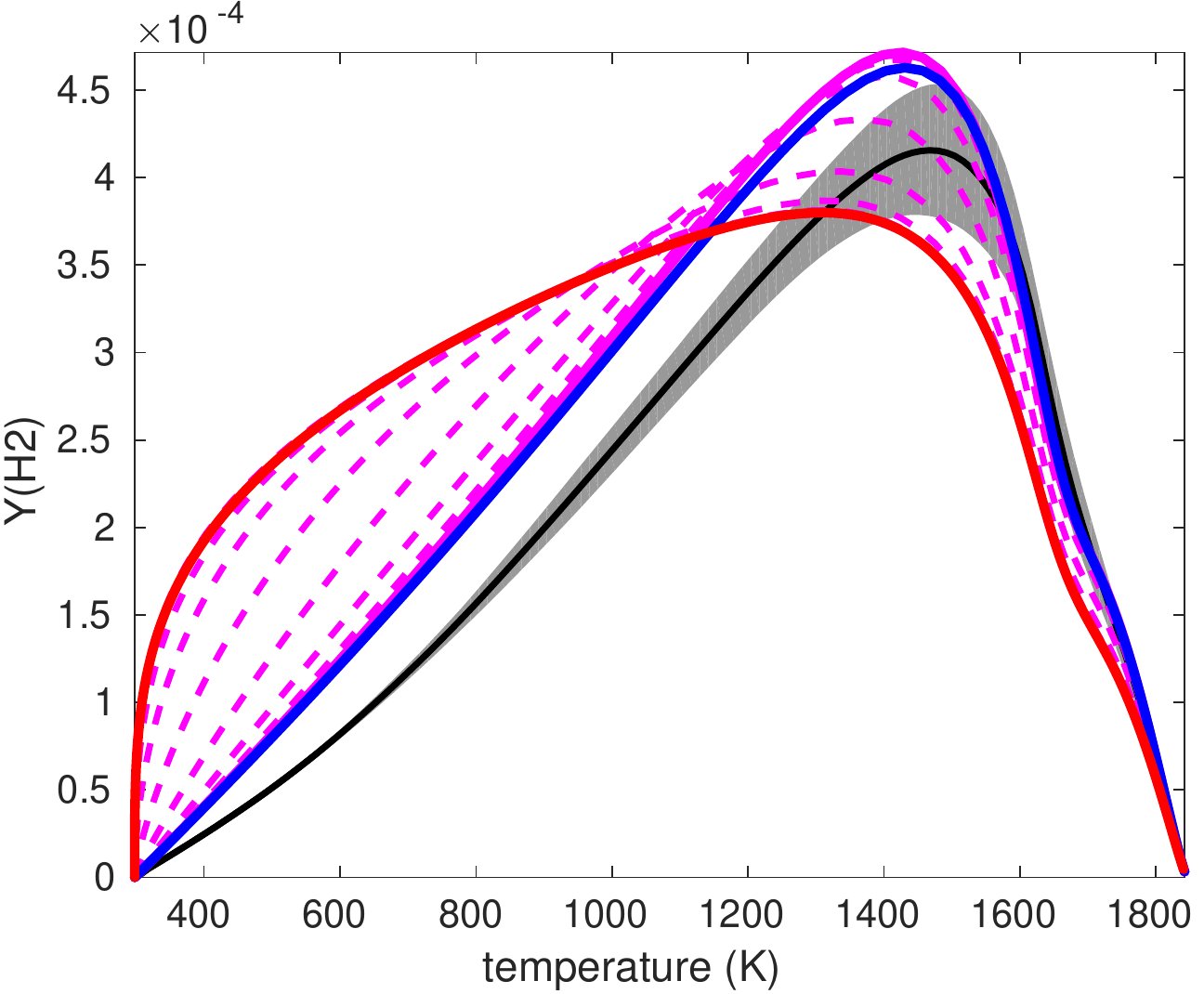}
\includegraphics[width=0.48\textwidth]{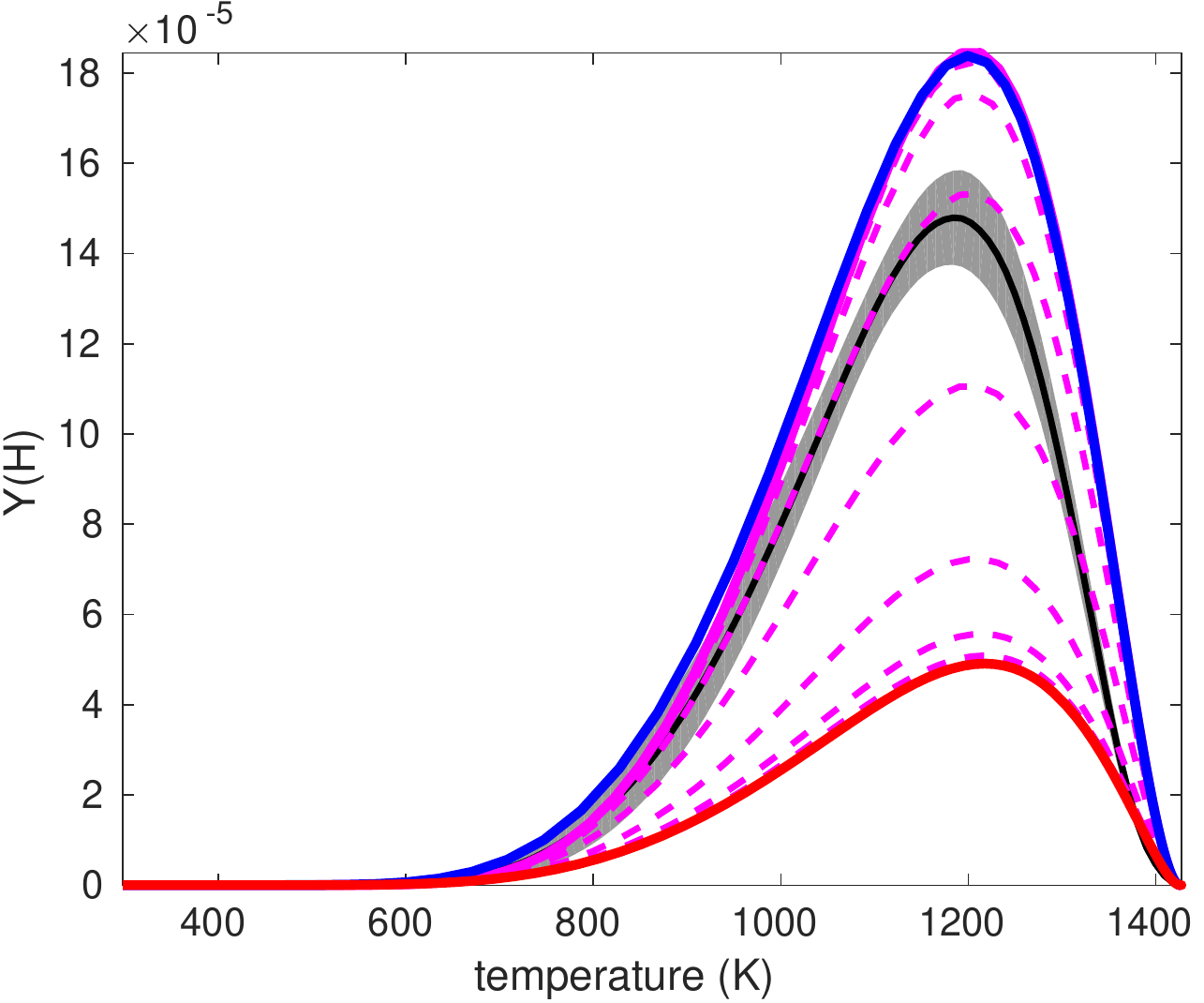}
\includegraphics[width=0.48\textwidth]{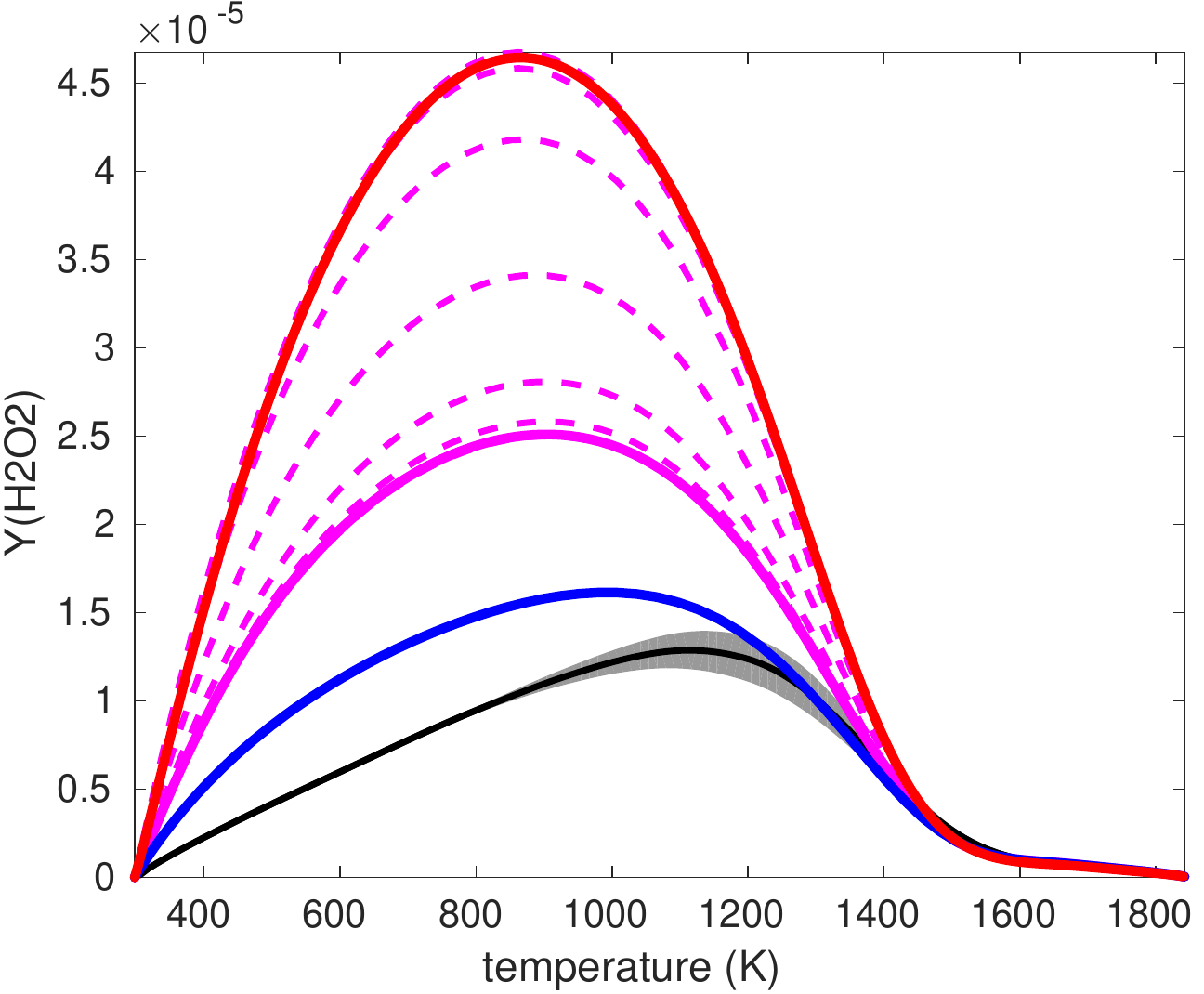}
\includegraphics[width=0.48\textwidth]{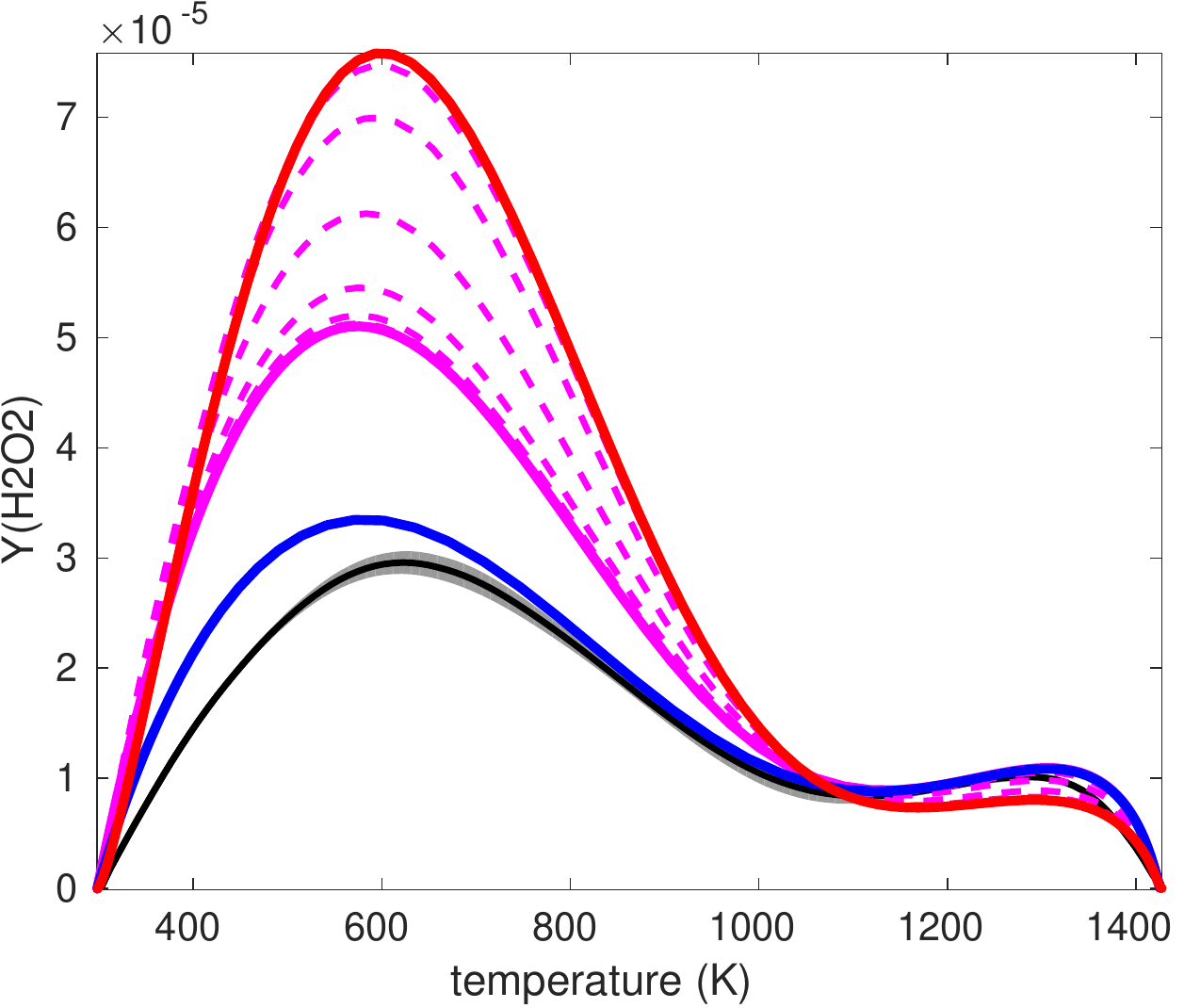}
\caption{Conditional means of species mass fraction for CH$_4$ (left) and H$_2$ (right) flames at $\Ka=8767$; see figure~\ref{fig:cm-fcr-hr} for a description of the line styles, with the addition of one standard deviation about the mean shown in grey.}
\label{fig:cm-y}
\end{figure}

For the methane flames, the turbulent response of conditional means of species mass fractions 
remain consistent with the classification set out in \citet{AspdenCnF16};
figure~\ref{fig:cm-y}(left column)
presents conditional means of mass fractions for CH$_4$, H$_2$ and H$_2$O$_2$
at $\Ka=8767$ (conditional means for all species and $\Ka$ are provided as
supplementary material).  
The fuel distribution is close to linear for all profiles, 
but there is the suggestion that the turbulent profile is slightly higher than the one-dimensional
profiles (not seen in the lower $\Ka$ cases).
The other two species present a strong response in the preheat
region; specifically, turbulence dominates species diffusion at high $\Ka$. The standard
deviation decreases with $\Ka$, and the distribution not only transitions from that comparable
with the laminar profile (red) towards the unity Lewis number profile (blue), but also appears
to reach a state that is distinct from all three of the one-dimensional profiles.
\citet{AspdenCnF16} suggested that a temperature-dependent diffusion coefficient may account
for the difference, and that it may be due the effect of dilatation on turbulence through the
flame; there may also be a change in chemical time scale.
Other species with distributions notably distinct are C$_2$H$_2$, HCCOH and CO
(see supplementary material).

A similar response is found for intermediate species in the hydrogen flames, but there
is a significantly different response in the fuel distribution; 
figure~\ref{fig:cm-y}(right column) presents conditional means of species mass fraction for H$_2$, H and H$_2$O$_2$ at $\Ka=8767$.
The fuel distribution has moved away from the (low $\Le$) one-dimensional laminar flame 
(that lies below the diagonal) towards the unity Lewis number profile that aligns with the 
diagonal mixing line.
The response of H is characteristic of O and OH (again, all species are presented as supplementary 
material), where the enhanced radical pool at low-to-moderate temperatures observed at lower $\Ka$ 
\citep[see][and supplementary material]{AspdenPCI15} is suppressed at $\Ka=8767$.
A strong response of H$_2$O$_2$ is observed in the preheat region, and again appears to tend
to a limit distinct from the three one-dimensional profiles.  
The standard deviation is also significantly reduced, almost to zero.

\section{Discussion and Conclusions}
\label{sec:conclusions}

Numerical simulations at extreme levels of turbulence
have shown a transition to distributed burning in lean premixed hydrogen flames.
The phenomenology of the transition is similar to that reported in an astrophysical context
\citep{Aspden08a}; there are no indications that suggest other fuels should not undergo a similar 
transition at sufficiently intense turbulence levels.
Unrealistically-high turbulence conditions were required to observe the transition, which is 
argued to be primarily due to dilatation and the higher density ratio between unburned and burned 
conditions (the density ratio is less than two for the supernova case, about four for hydrogen,
and over six for methane); this effect was not observed in constant-density reaction wave
propagation by \cite{YuPRE17}, for example. The effect of density ratio may also be compounded
by the thickness of the reaction zone relative to the thermal thickness; the reaction thickness
in the methane flame is relatively narrower than in the hydrogen flame, so turbulence has to
survive deeper into the methane flame to disrupt the reaction zone.

\begin{figure}
\centering
\includegraphics[width=0.48\textwidth]{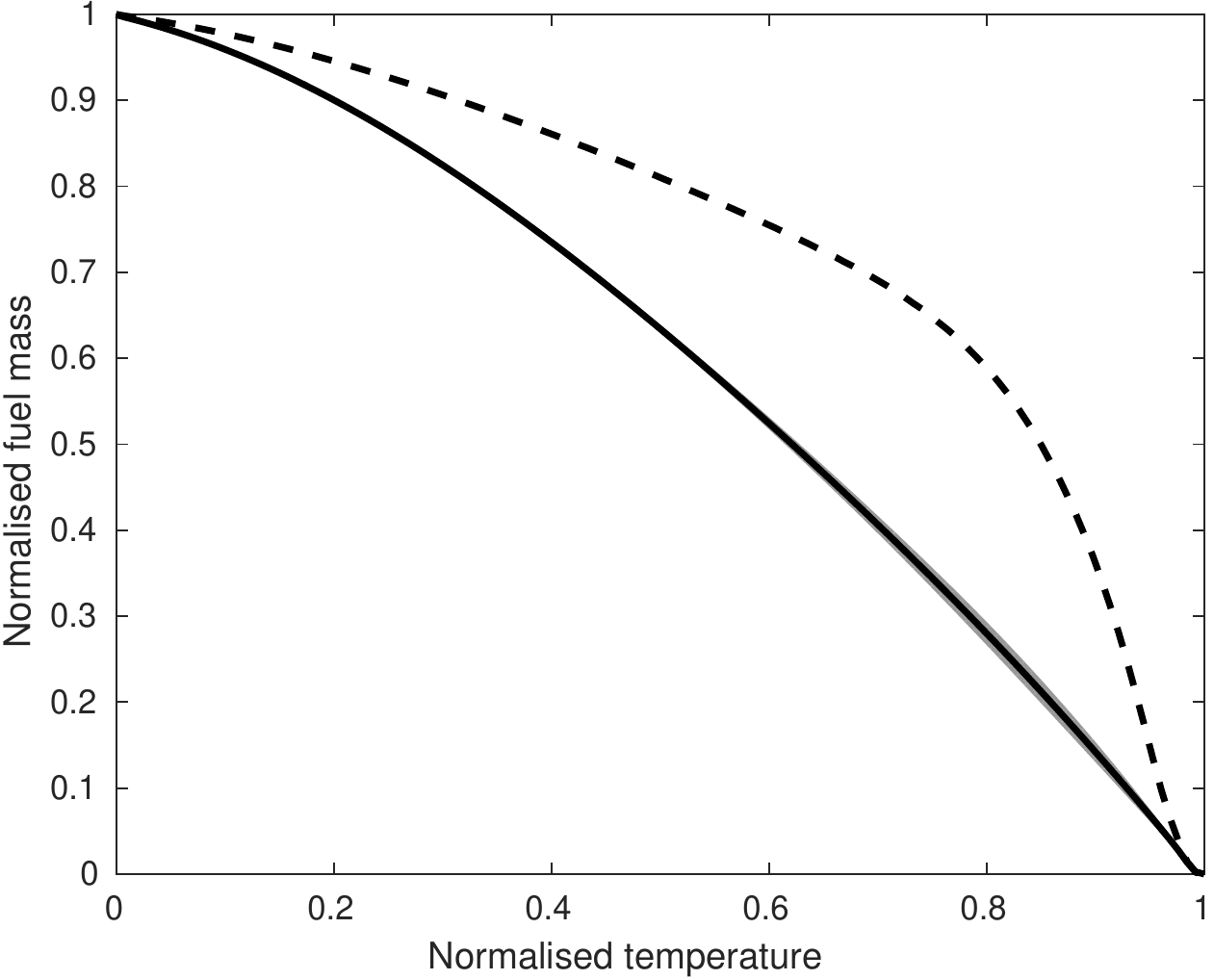}
\caption{Conditional mean of normalised fuel mass from the distributed supernova flame at $\Ka=266$ \citep{Aspden08a}; the conditional mean is the solid line with one standard deviation in grey, and the laminar flame profile is shown by the dashed line.}
\label{fig:cm-y-sne}
\end{figure}

The key behaviour characteristic of distributed burning not previously observed in chemical
flames is broadening of the reaction zone by turbulence; here, the heat release
region was visually found to be broadened by around an order of magnitude.  
Thermodiffusive effects (in particular, the super-adiabatic regions and the decorrelation of 
FCR and HR) were found to be suppressed, if not eliminated, with the peak
reaction rates falling from about fifteen times the peak laminar value to about four times.
Distributed reaction rates that are higher than the corresponding laminar flame is different from 
the response found in the supernova flame, where the peak reaction rates were found to be much
smaller.
This can be attributed to global Lewis number
by considering the fuel-temperature distribution, as discussed in \citet{AspdenPCI11b}.
Hydrogen has a low Lewis number, 
and so the fuel-temperature distribution of the laminar flame lies below the linear mixing 
distribution.  Strong turbulent mixing gives an effective unity Lewis number for all species 
(species and enthalpy are advected together in packets) so the distributed flame profile is 
close to linear -- fuel 
concentrations are higher at the same temperature, so reaction rates and heat release increase.
Conversely, for the supernova, the Lewis number is large, so the laminar flame profile lies
above the linear mixing distribution (see figure~\ref{fig:cm-y-sne}), 
and turbulent mixing results in fuel concentrations that
are lower at the same temperature, giving lower reaction rates.

Turbulent flame speeds were found to follow scaling laws at low-to-moderate Karlovitz numbers
(despite not satisfying the assumptions made), followed by an apparent transition. Even at the
extreme turbulence levels considered, there was insufficient evidence to draw solid conclusions about
behaviour in the distributed burning regime, and will require significant further work to demonstrate
the behaviour observed in the supernova flames \citep{Aspden10}.

The species distribution tends towards a limit that is distinct from the one-dimensional profiles
(laminar, unity Le, and turbulent diffusive limit), with a standard deviation of almost zero;
\cite{AspdenCnF16} suggested that a temperature-dependent diffusion coefficient is required to
account for dilatation through the flame.
A change in distribution was observed in particular for species that experience low-temperature 
activity (e.g.\ H$_2$O$_2$), which is argued to result from molecular diffusion of 
mobile species such as H$_2$ and/or H being overcome by turbulent mixing.

\subsection{Transition to distributed burning}

We continue to interpret the distributed burning regime as the limiting behaviour where turbulent
mixing dominates species and thermal diffusion to drive flame propagation, as discussed in
\cite{AspdenJFM11}, along with the consequences described therein.
It is well-established that turbulence increases the flame surface area and exaggerates the
thermodiffusive instability, but once turbulence is sufficiently intense, it can overcome dilatation
effects to disrupt the reaction zone.  We posit that the turbulently-mixed flame structure 
(i.e.~chemical distribution and reaction rates conditioned on some measure of progress variable
such as temperature) eventually reaches a limiting state (distinct from other one-dimensional
distributions considered here) that is invariant under further increases in turbulent intensity.  
The turbulent flame speed and thickness will respond to such increases in turbulent intensity 
(and so spatial gradients will eventually become small) but the underlying flame structure will 
not change (in temperature space).

We argue that only turbulence-flame interactions around the flame scales,
i.e.~length scales comparable with the thermal and reaction thicknesses,
are relevant to the transition; larger scales serve to move the turbulent
flame structure around, and smaller scales are not dynamically important.  Therefore, the
relevant dimensionless quantity necessary for distributed burning is sufficiently high
Karlovitz number, defined according to equation~(\ref{eq:Ka}), because it appropriately 
characterises the turbulence in terms of the inertial subrange.  
As Karlovitz number increases, turbulence is able to penetrate further into the flame,
until eventually it is sufficiently strong enough to overcome dilatation effects and 
mix the reaction zone quicker than it can burn.
While these simulations have a particularly small integral length scale, 
they are representative of how turbulence at the same Karlovitz number would interact with the
flame; the response to increasing the integral length scales at a fixed Karlovitz number in the
distributed regime was demonstrated in \cite{Aspden10}, consistent with \citet{Zimont79}.

The critical Karlovitz number necessary is fuel-dependent, and will be higher where the density 
jump across the flame is large (due to suppression of turbulence by dilatation); it is not 
universal, and it is certainly not 100.  The global Lewis number (that of the deficient species) 
will also affect this critical Karlovitz number.
The transition to distributed burning increases the local reaction rates for low $\Le$ 
(and vice versa for high $\Le$), which suggests that the critical Karlovitz number required to 
transition will be lower at high $\Le$; if the reaction rates decrease, then there is more time 
for turbulence to interact with the reaction zone, allowing it to mix before it can burn.
Moreover, the Karlovitz number defined in equation~(\ref{eq:Ka}) is based on the laminar
flame properties (or indeed freely-propagating values in the case of hydrogen), but in a 
distributed flame, the global Lewis number becomes irrelevant because molecular diffusion is
dominated by turbulent mixing; if one were to artificially alter the global Lewis number, the 
reference flame properties would be different, but the corresponding distributed flames should be 
anticipated to be identical for the same turbulence conditions (not the same $\Ka$).  
This suggests that the critical Karlovitz number could be evaluated based on a one-dimensional
calculation that somehow accounts for the subtle interaction of an appropriate turbulent mixing 
model (with an effectively unity Lewis number) and the resulting chemical time scale.
Furthermore, the use of the freely-propagating reference values may only be relevant for the thin
reaction zone.

We contend that the present hydrogen flame at the highest $\Ka$ has indeed achieved the distributed
burning regime.  Note that the limiting case is not confirmed definitively by the present 
simulations as that would require yet higher $\Ka$ to establish invariant behaviour (although
the $\Ka=974$ case is not substantially different). 
What is confirmed by the present simulations is a change in flame structure,
especially a spatial broadening of the reaction zone by turbulence, accompanied by a
($\Le$-dependent) change in magnitude, and a distinct chemical distribution.

\subsection{Potential for realising distributed burning}\label{sec:cantera}

\begin{figure}
\centering
\makebox[0.5\textwidth][c]{Methane flames}\makebox[0.5\textwidth][c]{Hydrogen flames}
\includegraphics[width=0.48\textwidth]{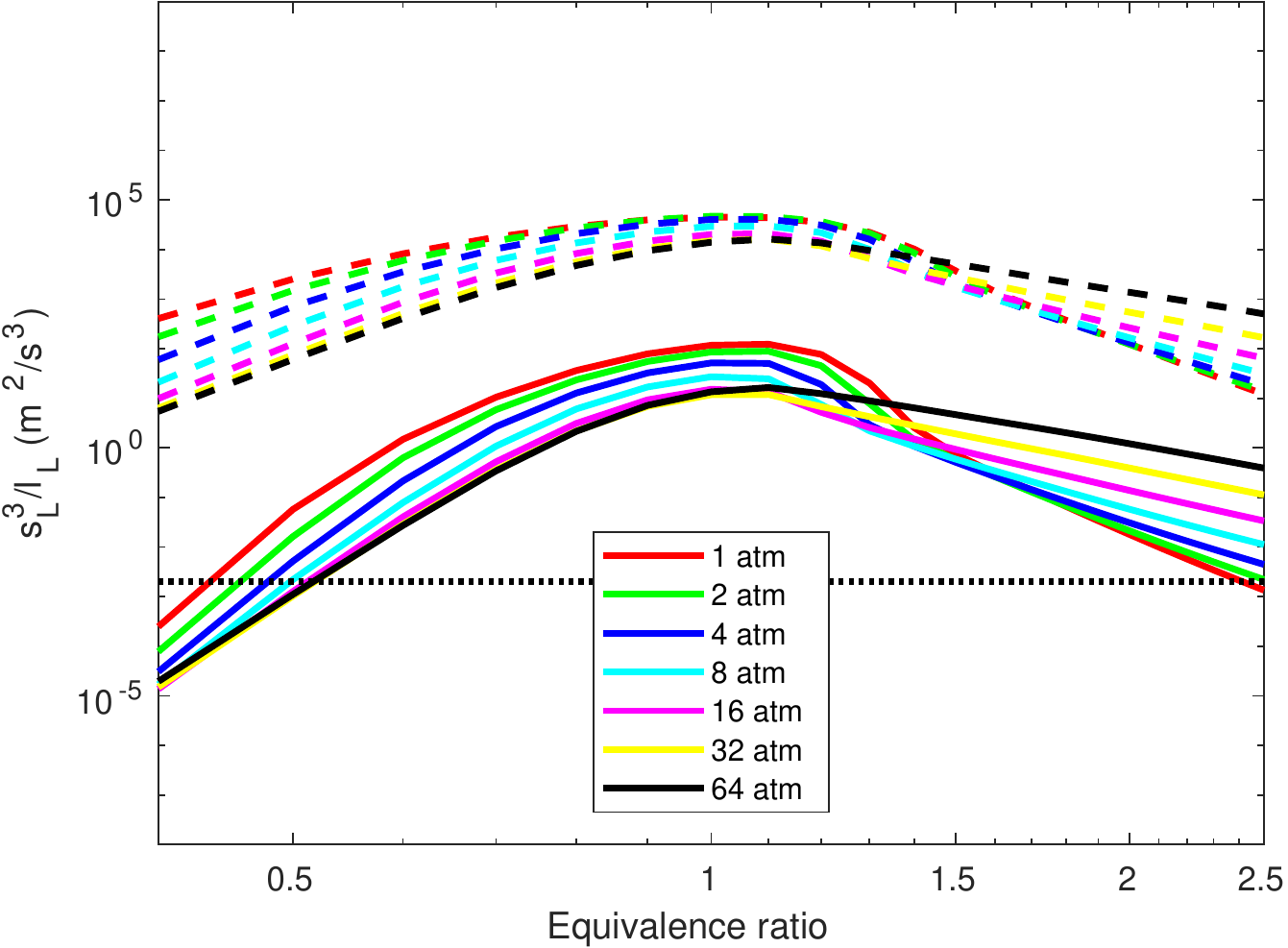}
\includegraphics[width=0.48\textwidth]{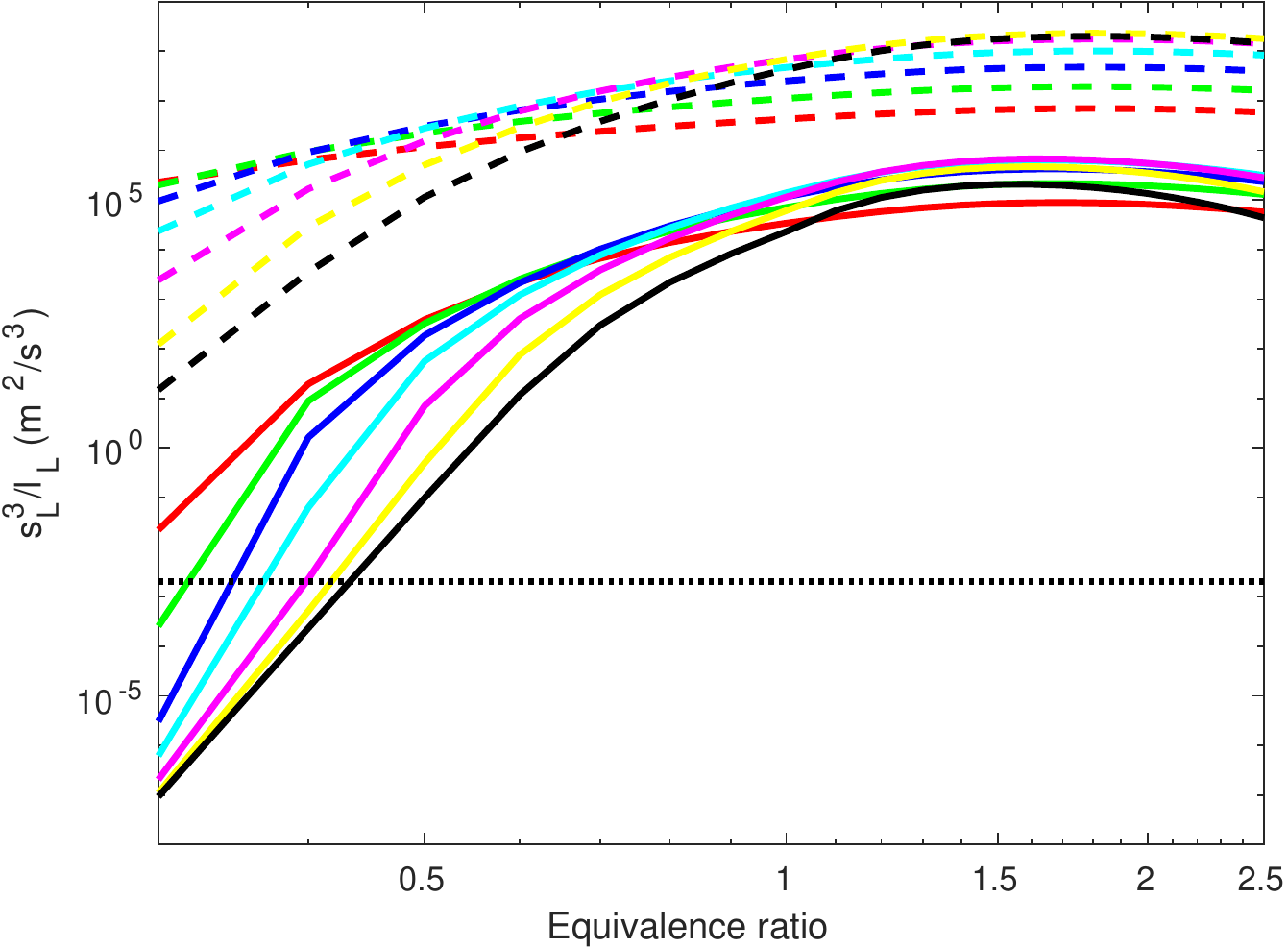}
\caption{Laminar flame properties from methane (left) and hydrogen (right) flames; solid and
  dashed lines denote reactant temperatures of 300\,K and 800\,K, respectively.
  Maximising Karlovitz number is equivalent to minimising $s_L^3/l_L$; given sample conditions
  of $u^\prime=10$\,m/s at $l=5$\,mm, a Karlovitz number in excess of 10,000 can be achieved
  for reactant conditions with $s_L^3/l_L$ below 0.002\,m$^2$/s$^3$, as shown by the dotted horizontal line.}
\label{fig:cantera}
\end{figure}

Whether or not there is a practical use for distributed flames remains to be seen, but it remains
a phenomenon of academic interest, in particular as the limiting case of high-Karlovitz-number
turbulence-flame interaction.
The conditions required here for distributed burning are unphysical and unrealisable in practice, 
but there are steps that could be taken that may lead to distributed burning.
To explore potential conditions, a series of one-dimensional laminar flame calculations were run 
using Cantera \citep{cantera} to obtain the laminar flame speed ($s_L$), the thermal thickness 
($l_L$) and the density ratio ($P=\rho_u/\rho_b$) between reactant and products.
Reactant temperatures were varied between 300\,K and 800\,K, pressures from 1\,atm to 64\,atm,
and equivalence ratios from 0.3 for H$_2$ and 0.4 for CH$_4$ (and C$_{12}$H$_{26}$; 
see supplementary material) up to 2.5.

Since the strength of the turbulence that interacts at the flame scale depends solely
on $\Ka$, then for given turbulence conditions $u^\prime$ and $l$, finding reactant conditions
favourable for distributed burning (i.e.~maximising $\Ka$) is equivalent to minimising
$s_L^3/l_L$.  Starting with example realisable conditions,
say $u^\prime=10$\,m/s at $l=5$\,mm, then obtaining a Karlovitz number of 10,000 requires the ratio
$s_L^3/l_L$ to be below 0.002\,m$^2$/s$^3$.
This ratio is shown for methane and hydrogen flames as a function
of equivalence ratio at $T=300$\,K (solid lines) and 800\,K (dashed lines) for all pressures
(all other data is presented as supplementary material).  The figure shows that the likelihood
of realising distributed burning can be increased by increasing pressure, decreasing equivalence
ratio, and decreasing reactant temperature.  Preheating reduces the density ratio, and therefore
the impact of dilatation on turbulence through the flame, but increasing the reactant temperature
from 300\,K to 800\,K only reduces the density ratio by a factor of about two.  The increase in
flame speed is closer to two orders of magnitude, and so is likely to far outweigh the benefit
of reduced density ratio; preheating is unlikely to be favourable for the 
transition to distributed burning.

\vspace{2mm}

MSD and JBB were supported by the DOE Applied Mathematics Research Program of the 
DOE Office of Advanced Scientific Computing Research under the U.S. Department of Energy Contract 
No.~DE-AC02-05CH11231, which included computational resources at the National Energy Research 
Scientific Computing Center (NERSC).\\

\vspace{2mm}

\appendix

\section*{Appendix A}\label{appA}

\begin{figure}
\centering
\includegraphics[width=0.48\textwidth]{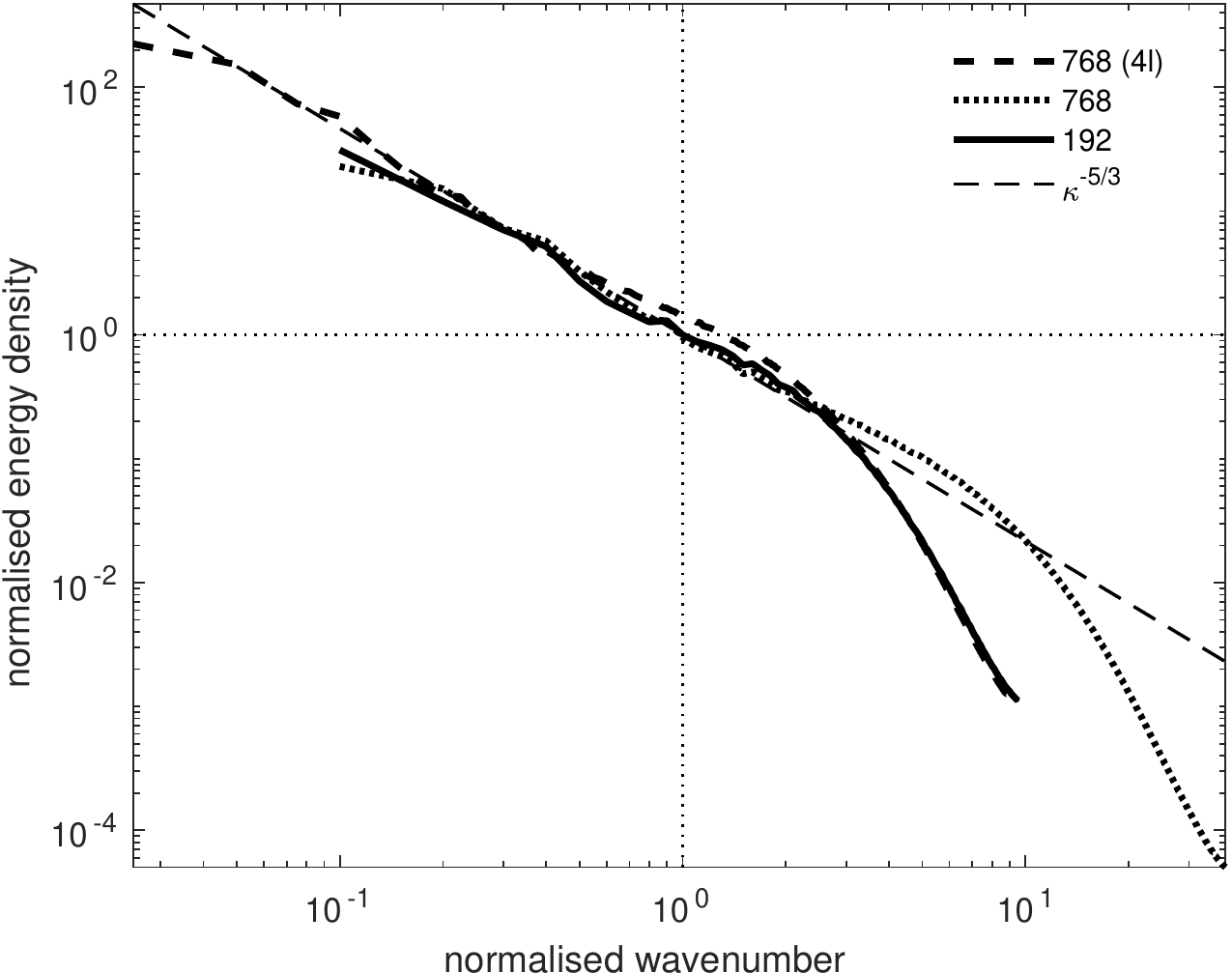}
\includegraphics[width=0.48\textwidth]{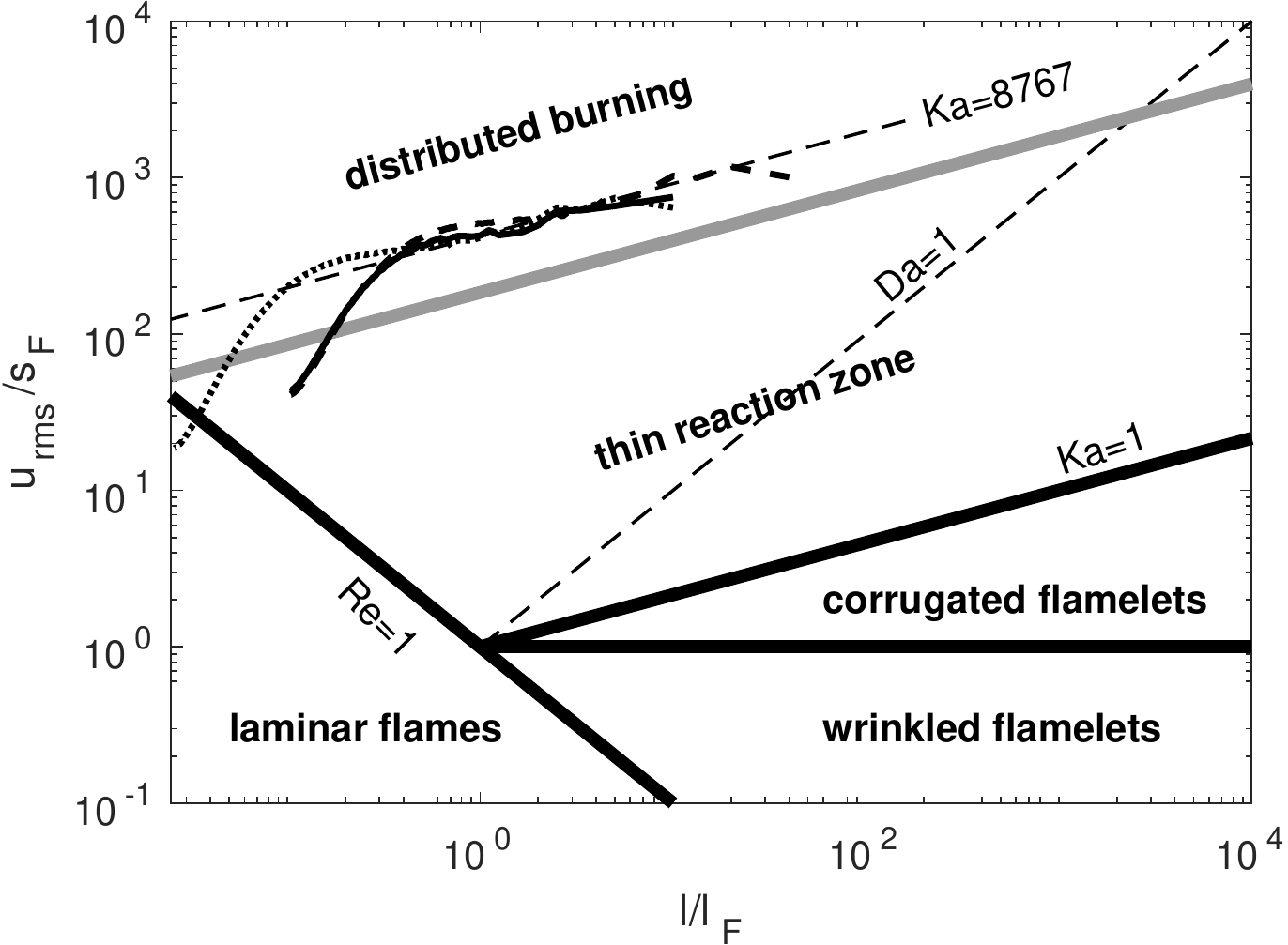}
\caption{Normalised kinetic energy density spectrum as a function of wavenumber normalised by the thermal thickness of the laminar flame (left), and the same recast schematically on the regime diagram (right).}
\label{fig:spectrum}
\end{figure}

To explore the potential consequences of the lack of resolution on the turbulence,
this appendix presents simulations of homogeneous isotropic turbulence in a triply-periodic cube
at the same (reactant) conditions as the hydrogen flame case at the highest Karlovitz number.
One simulation was run at $192^3$, corresponding directly to the same resolution as the flame
simulation, another at $768^3$ with the same integral length scale and therefore four times the 
resolution, and a final case at $768^3$ with an integral length scale four times greater (denoted 
$4l$) and therefore matching the cell size; in this last case, the magnitude of the forcing was 
increased to match the energy dissipation rate, which corresponds to matching the Karlovitz number.

The kinetic energy density spectra for the three cases are shown in figure~\ref{fig:spectrum} (left),
where the wavenumber has been normalised by the laminar flame thickness and the energy density 
has been normalised by that at the flame scale in the $192^3$ case.  Note the coincidence of the
inertial subranges, however short, and that they closely follow the expected five-thirds cascade.
The inertial subrange of the $192^3$ case is clearly truncated by the lack of small-scale resolution,
however this occurs at wavenumbers greater than the flame scale, which supports the argument
that the scales represented on the grid are adequate to capture turbulence-flame interactions.
The energy density spectra have also been recast schematically on the regime diagram
by noting $\sqrt{\kappa E}$ is representative of the velocity scale,
where $\kappa$ and $E$ are wavenumber and energy density, respectively.
This illustrates how the inertial subrange follows a constant Karlovitz line (shown by the dashed 
black line; referred to as turbulence lines by \citet{Poinsot05}, for example) because the energy 
dissipation rate is constant throughout ($\Ka\sim\varepsilon^{1/2}\sim\,$const), and where the lack 
of resolution compromises the calculation 
(the inertial range should continue to follow the constant Karlovitz line), again lending support
to the argument that the lack of resolution will not be detrimental to resolving the flame physics.
It is this coincidence of the inertial subrange with the constant Karlovitz line that supports the 
use of so-called small-eddy simulation approach for turbulent-flame interactions, and for the use of
the present definition of Ka (rather than one involving Kolmogorov scales or viscosity); 
specifically, it is the appropriate definition characterising turbulence-flame interactions,
not one based on the Kolmogorov time scale and the assumption that $\nu\approx s_Fl_F$.
The simulations presented here are underresolved, which leads to a smaller effective Reynolds number
but the Karlovitz number (appropriately defined) is the same.

\section*{Appendix B}\label{appB}

\begin{figure}
\centering
\iflowres
\includegraphics[width=0.48\textwidth]{conv/plt42963-lr}
\includegraphics[height=0.3\textwidth]{conv/plt42963-CH2O-overlay+-lr}
\else
\includegraphics[width=0.48\textwidth]{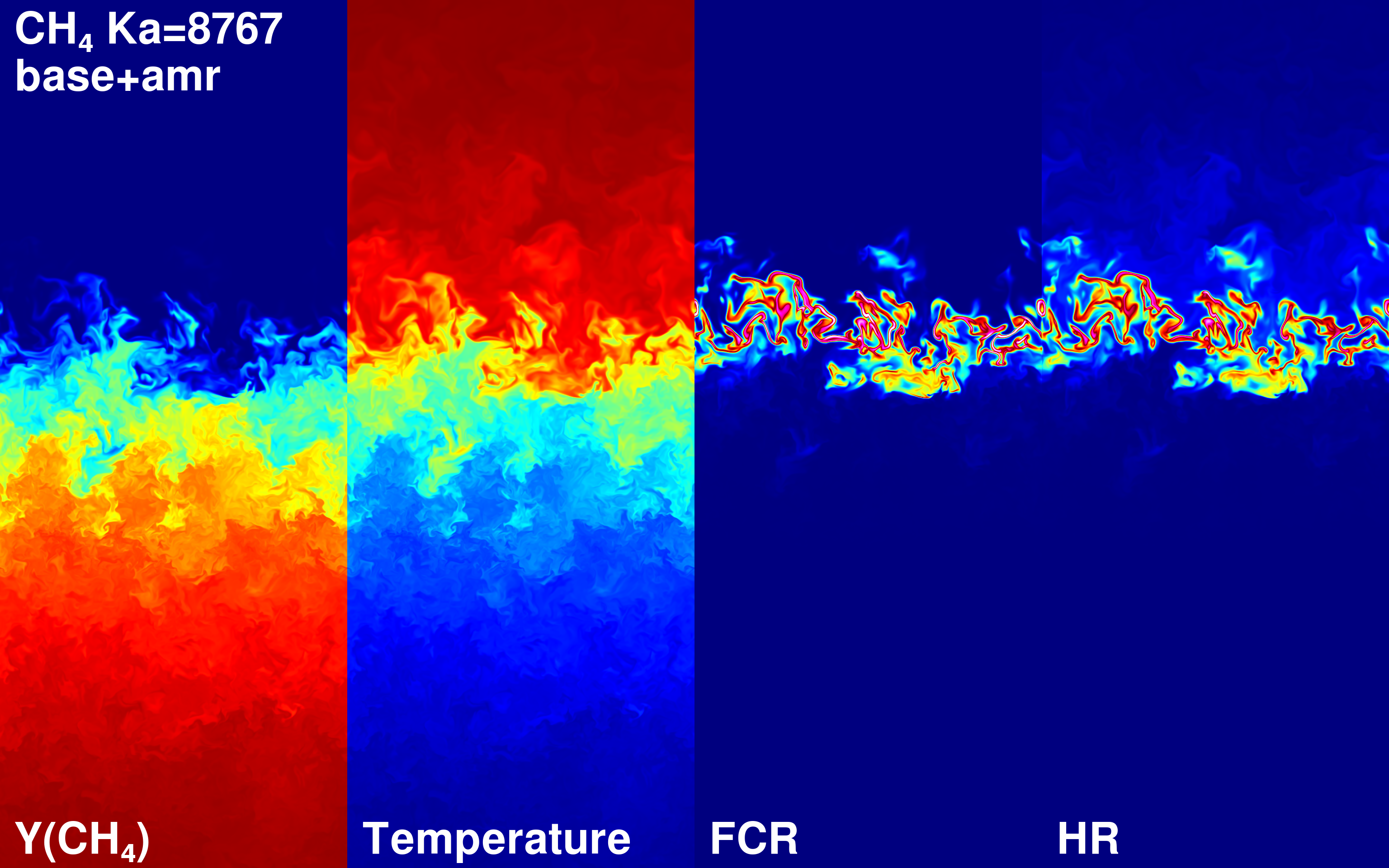}
\includegraphics[height=0.3\textwidth]{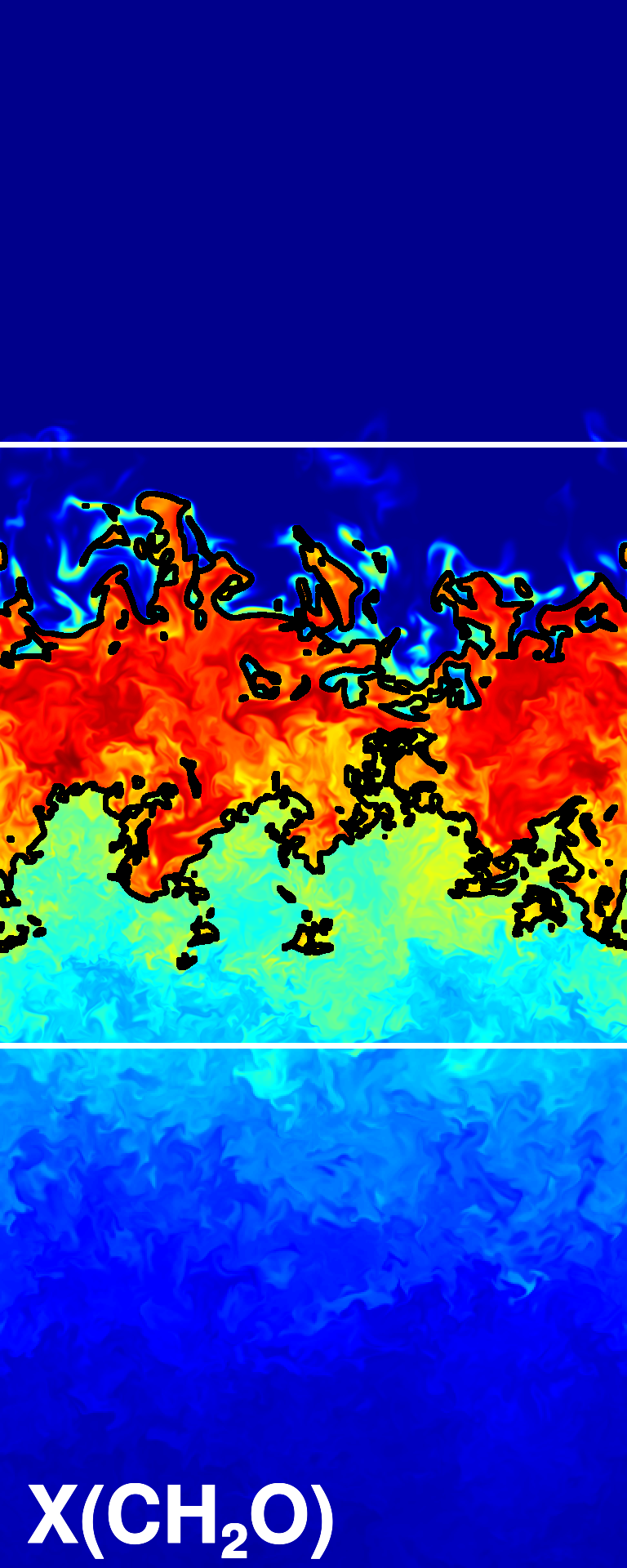}
\fi
\caption{Slices of fuel mass fraction, temperature, fuel consumption rate, and heat release for the CH$_4$ flame at $\Ka=8767$ run with adaptive mesh refinement (for comparison with figure~\ref{fig:slices}), along with formaldehyde mole fraction, which was used for gridding criterion (the black contour plus a 16-cell buffer resulted in the region between the two horizontal white lines being refined, capturing the reaction layer and a substantial region ahead of it).}
\label{fig:slicesConv}
\end{figure}

\begin{figure}
\centering
\includegraphics[width=0.48\textwidth]{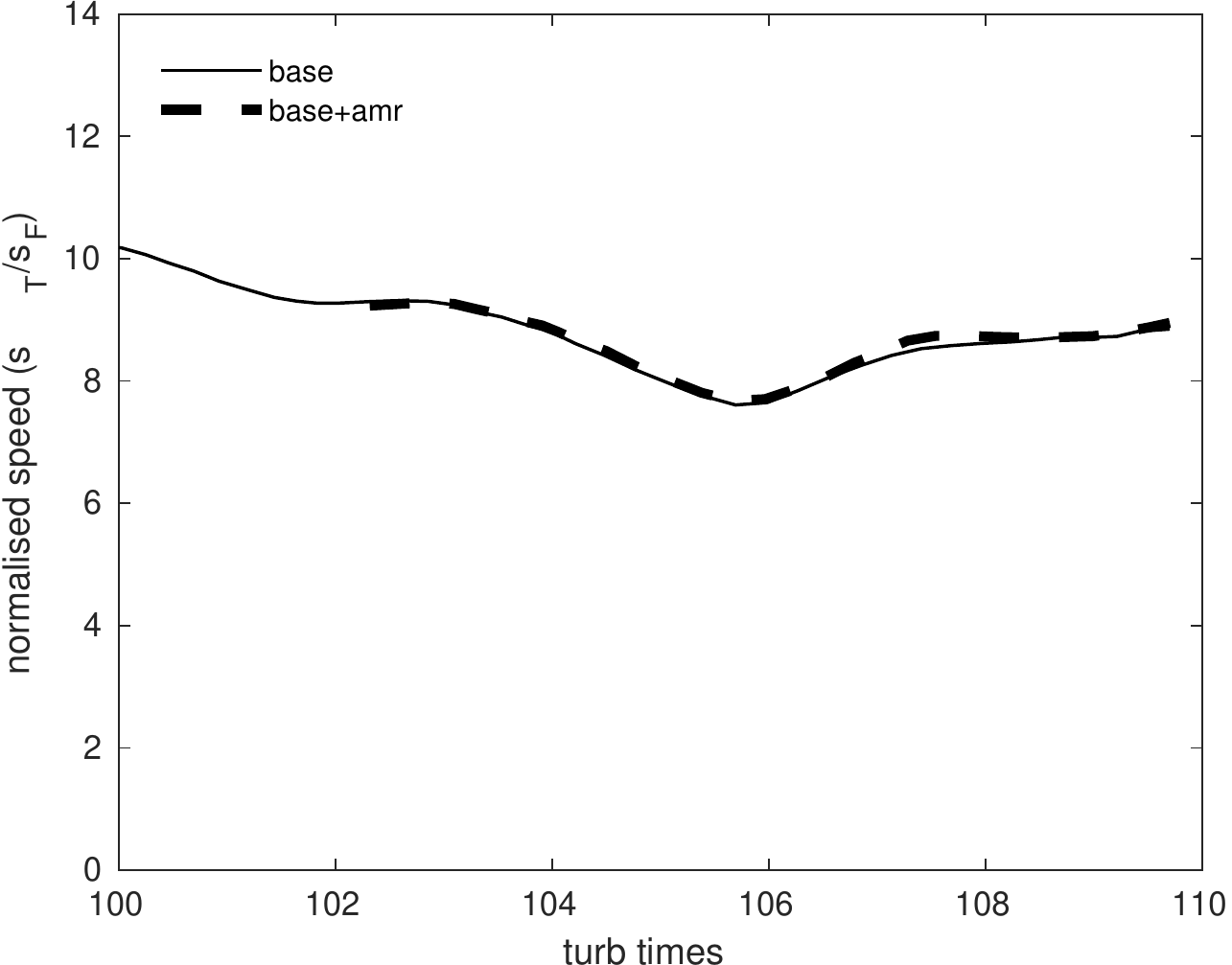}
\includegraphics[width=0.48\textwidth]{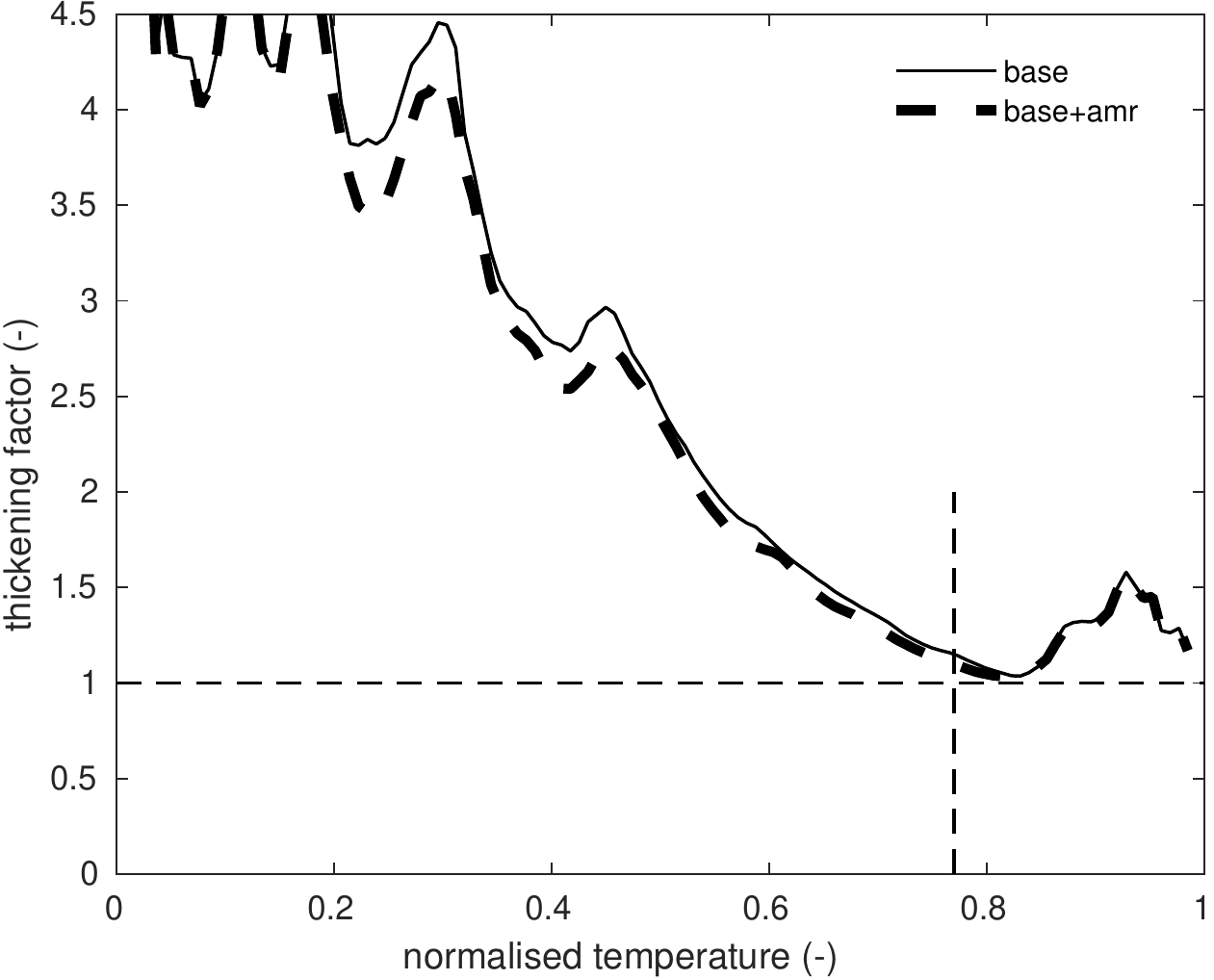}
\includegraphics[width=0.48\textwidth]{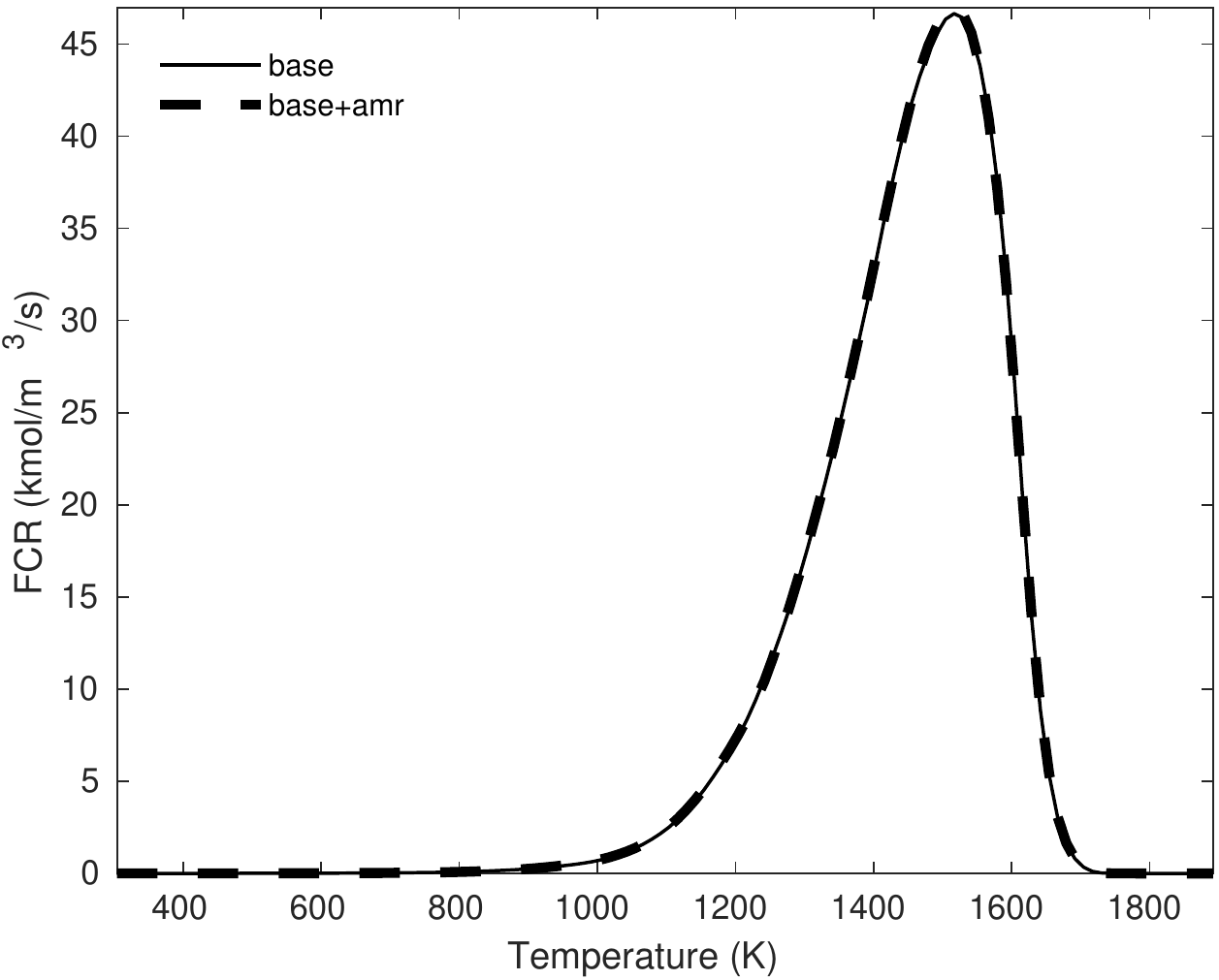}
\includegraphics[width=0.48\textwidth]{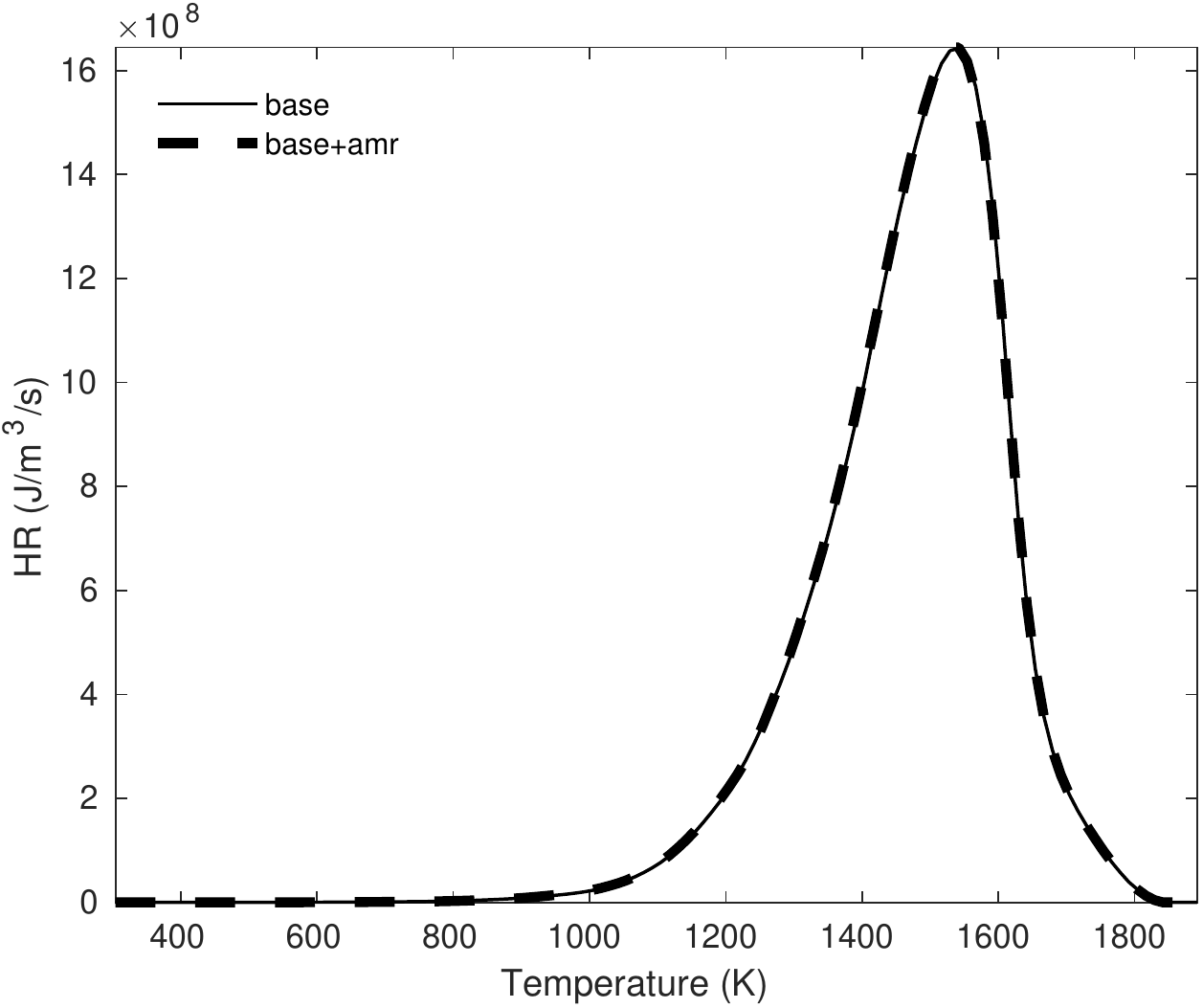}
\caption{Comparison of turbulent flame speeds, thickening factor, and conditional means of fuel consumption rate and heat release with adaptive mesh refinement for the CH$_4$ flame at $\Ka=8767$.}
\label{fig:convCH4}
\end{figure}

To establish the consequences for not completely resolving the spatial scales down to the Kolmogorov 
length scale in the methane flame at the highest Karlovitz number, this appendix
presents an additional simulation with higher resolution; this was achieved by restarting
the calculation with one level of adaptive mesh refinement around the flame.  Refinement was
added based on formaldehyde mole fraction (above $5\times10^{-4}$) with an additional buffer of 16
computational cells, following \cite{AspdenJFM11} to allow the higher wavenumbers to become
populated; at this higher resolution there are over 38 computational cells across the 
thermal thickness of the laminar flame.

Figure~\ref{fig:slicesConv} shows slices of fuel, temperature, fuel consumption rate and heat
release for the case with adaptive mesh refinement (for comparison with figure~\ref{fig:slices}),
along with formaldehyde mole fraction with the gridding threshold contour shown in black and
the resulting region of increase resolution shown by the two white lines.
While it is unreasonable to expect that the turbulence structure is exactly the same, 
the general flame features appear unaffected by the increase in resolution.

Figure~\ref{fig:convCH4} shows normalised turbulent flame speed, thickening factor and
conditional means of fuel consumption rate and heat release (conditioned on temperature);
in each case, the base grid is shown as a thin solid line, and the case with adaptive mesh
refinement is shown as a thick dashed line.  The temporal averaging was performed over the final
two-thirds of the restarted run (corresponding to about five eddy turnover times), 
and the same time period was also used at the lower resolution.
The turbulent flame speed (a global metric) agrees almost exactly between resolutions, 
and the conditional means of fuel consumption rate and heat release (local metrics) are 
indistinguishable.  There do appear to be differences in the thickening factor between the 
two resolutions, specifically, the thickening factor is lower
at the higher resolution, indicative of slightly higher gradients;
both the higher gradient and disparity are not particularly surprising, especially for a
higher moment in a short period of temporal averaging.
Ultimately, this higher resolution simulation demonstrates that the effect of the apparent lack of 
resolution is small.

\bibliographystyle{plainnat}
\bibliography{paper}

\end{document}